\documentclass[aps,pre,onecolumn,notitlepage,groupedaddress,superscriptaddress,nofootinbib,longbibliography]{revtex4-1}

\usepackage{amsmath,amsbsy,amssymb, graphicx, textcomp, upgreek, hyperref, subfigure, dsfont, accents}

\usepackage[dvipsnames]{xcolor} % must be defined before tikz and todonotes are loaded

\usepackage{tikz}
\usetikzlibrary{calc}
\usetikzlibrary{fpu}
\usetikzlibrary{arrows}
\usetikzlibrary{decorations,decorations.shapes}

\makeatletter
\newcommand*{\balancecolsandclearpage}{%
  \close@column@grid
  \clearpage
  \twocolumngrid
}
\makeatother
\begin{document}

%Title of paper
\title{Thermophoresis of Janus particles at large Knudsen numbers}

\author{Tobias Baier}
%\email[]{baier@nmf.tu-darmstadt.de}
\affiliation{Institute for Nano- and Microfluidics, Center of Smart Interfaces, Technische Universit\"at Darmstadt, Germany}

\author{Sudarshan Tiwari}
%\email[]{tiwari@mathematik.uni-kl.de}
\affiliation{Department of Mathematics, Technische Universit\"at Kaiserslautern, Germany}

\author{Samir Shrestha}
%\email[]{samirstha@ku.edu.np}
\affiliation{Department of Natural Sciences, Kathmandu University, Nepal}

\author{Axel Klar}
%\email[]{klar@mathematik.uni-kl.de}
\affiliation{Department of Mathematics, Technische Universit\"at Kaiserslautern, Germany}

\author{Steffen Hardt}
%\email[]{hardt@nmf.tu-darmstadt.de}
\affiliation{Institute for Nano- and Microfluidics, Center of Smart Interfaces, Technische Universit\"at Darmstadt, Germany}

%\date{\today}

\begin{abstract}
The force and torque on a Janus sphere moving in a rarefied gas with a thermal gradient are calculated. The regime of large Knudsen number is considered, with the momenta of impinging gas molecules either obtained from a Chapman-Enskog distribution or from a binary Maxwellian distribution between two opposing parallel plates at different temperature. The reflection properties at the surface of the Janus particle are characterized by accommodation coefficients having constant but dissimilar values on each hemisphere. It is shown that the Janus particle preferentially orients such that the hemisphere with a larger accommodation coefficient points towards the lower temperature. The thermophoretic velocity of the particle is computed, and the influence of the thermophoretic motion on the magnitude of the torque responsible for the particle orientation is studied. The analytical calculations are supported by Direct Simulation Monte Carlo results, extending the scope of the study towards smaller Knudsen numbers. The results shed light on the efficiency of oriented deposition of nanoparticles from the gas phase onto a cold surface.
\end{abstract}

% insert suggested PACS numbers in braces on next line
%\pacs{47.61.-k, 47.45.Dt, 05.60.-k, 44.15.+a}

% insert suggested keywords - APS authors don't need to do this
%\keywords{}

%\maketitle must follow title, authors, abstract, \pacs, and \keywords
\maketitle

%%%%%%%%%%%%%%%%%%%%%%%%%%%%%%%%%%%%%%%%%%%%%%%%%%%%%%%%
\section{Introduction\label{sec:intro}}

\begin{figure}
%	\includegraphics{Fig1}
%%%%%%%%%%%%%%%  tikzpicture  %%%%%%%%%%%%%%%%%%
	\raisebox{-0.5\height}{
		\begin{tikzpicture}[scale=1, >=stealth]
		  Local definitions
		\def\angTheta{55}
		\def\lenR{1} % radius
		\def\lenH{2} % gap width
		\def\lenW{2} % plot width
		
		% define vertices
		\coordinate (p1) at ($(-\lenW,\lenH)$);
		\coordinate (p2) at ($(\lenW,\lenH)$);
		\coordinate (p3) at ($(-\lenW,-\lenH)$);
		\coordinate (p4) at ($(\lenW,-\lenH)$);
		
		\coordinate (p5) at ($(-\lenW,0)$);
		\coordinate (p6) at ($(\lenW,0)$);
		
		\coordinate (c0) at (0,0);
		\coordinate (t1) at ($(90:\lenR)$);
		\coordinate (t2) at ($(90-\angTheta:\lenR)$);
		\coordinate (t3) at ($(180-\angTheta:\lenR)$);
		\coordinate (t4) at ($(0-\angTheta:\lenR)$);
		
		% draw lines   
		\draw[very thick,color=Black] (p1) -- node[black,above] {$T_h$} (p2);		% top line
		\draw[very thick,color=Black] (p3) -- node[black,below] {$T_l$} (p4);		% bottom line
		\draw[dashed] (t3) -- (t4);												% center line
		
		\draw[very thick, violet, fill=violet!10] (t4) arc (-\angTheta:180-\angTheta:1) -- (t4);
		\draw[very thick,color=Blue, fill=blue!10] (t3) arc (180-\angTheta:360-\angTheta:1);		% circle
		\draw[->, very thick] ($(c0)$) -- node[black,below right] {$\mathbf{n}_p$} ($0.75*(t2)$);
		\draw[->, very thick] ($(c0)$) -- node[black,right=4pt] {$\mathbf{n}_p$} ($0.75*(t2)$);
		\draw[dashed] (c0) -- (t1);
		
		% \draw[->, very thick,color=BrickRed] ($(-3/4*\lenW,0)$) -- node[black,left] {$\mathbf{e}_z$} ($((-3/4*\lenW,\lenR*3/4)$);
		 \draw[->, very thick,color=BrickRed] (t2) -- node[black,left] {$\mathbf{e}_r$} ($(t2)+(90-\angTheta:\lenR*3/4)$);
		 \draw[->, very thick,color=BrickRed] (t2) -- node[black,above] {$\mathbf{e}_\vartheta$} ($(t2)+(-\angTheta:\lenR*3/4)$);
		
		% \pgfmathparse{90-\angTheta)}  																	% calculate arc angle for alpha
		\draw ($(c0)+(90:0.5)$) arc (90:90-\angTheta:0.5) ;
		\draw ($(c0)+(90-\angTheta/2:\lenR*2/3)$) node {$\theta$};
		
		\draw[->, very thick,color=BrickRed] ($(-\lenW,0)$) -- node[black,right] {$\mathbf{e}_z$} ($((-\lenW,\lenR*3/4)$);
		\draw[->, very thick,color=BrickRed] ($(-\lenW,0)$) -- node[black,below] {$\mathbf{e}_x$} ($((-\lenW+\lenR*3/4,0)$);

		 draw plate distance
		\def\lenOffset{0.5} % offset
		\coordinate (a1) at ($(\lenW+\lenOffset,-\lenH)$);
		\coordinate (a2) at ($(\lenW+\lenOffset,\lenH)$);
		\draw[<->, very thick] (a1) -- node[black,right] {$h$} (a2);
		 draw diameter
		\def\lenOffset{-0.25} % offset
		\coordinate (b1) at ($(\lenW+\lenOffset,-\lenR)$);
		\coordinate (b2) at ($(\lenW+\lenOffset,\lenR)$);
		\draw[<->, very thick] (b1) -- node[black,right] {$2R$} (b2);
		
		\end{tikzpicture}
	}
	%
	%%%%%%%%%%%%%%%%%%%%%%%%%%%%%%%%%
	\hspace{1.5cm}
	%%%%%%%%%%%%%%%%%%%%%%%%%%%%%%%%%
	%
	\raisebox{-0.5\height}{	
		\begin{tikzpicture}[scale=1, >=stealth]
		  Local definitions
		\def\angTheta{40}
		\def\lenR{1} % radius
		\def\lenH{2} % gap width
		\def\lenW{2} % plot width
		
		% define vertices
		\coordinate (p1) at ($(-\lenW,\lenH)$);
		\coordinate (p2) at ($(\lenW,\lenH)$);
		\coordinate (p3) at ($(-\lenW,-\lenH)$);
		\coordinate (p4) at ($(\lenW,-\lenH)$);
		
		\coordinate (p5) at ($(-\lenW,0)$);
		\coordinate (p6) at ($(\lenW,0)$);
		
		\coordinate (c0) at (0,0);
		\coordinate (t1) at ($(90:\lenR)$);
		\coordinate (t2) at ($(90-\angTheta:\lenR)$);
		
		% draw lines   
			\draw[very thick,color=Black] (p1) -- node[black,above] {$T_h$} (p2);		% top line
			\draw[very thick,color=Black] (p3) -- node[black,below] {$T_l$} (p4);		% bottom line
			\draw[very thick,gray, dashed] (p5) -- (p6);												% center line
		
		\draw[very thick,color=Blue] (t1) arc [start angle=90, end angle = 450, radius = 1];		% circle
		\draw[->, very thick] ($(c0)$) -- node[black,right] {$\mathbf{r}$} ($(t2)$);
		\draw[dashed] (c0) -- (t1);
		
		\draw[->, very thick,color=BrickRed] (t2) -- node[black,left] {$\mathbf{e}_r$} ($(t2)+(90-\angTheta:\lenR*3/4)$);
		\draw[->, very thick,color=BrickRed] (t2) -- node[black,above] {$\mathbf{e}_\vartheta$} ($(t2)+(-\angTheta:\lenR*3/4)$);
		
		\draw[->, very thick,color=BrickRed] ($(-\lenW,0)$) -- node[black,right] {$\mathbf{e}_z$} ($((-\lenW,\lenR*3/4)$);
		\draw[->, very thick,color=BrickRed] ($(-\lenW,0)$) -- node[black,below] {$\mathbf{e}_x$} ($((-\lenW+\lenR*3/4,0)$);
		
		\pgfmathparse{90-\angTheta)}  																	% calculate arc angle for alpha
		\draw ($(c0)+(90:0.5)$) arc (90:55:0.5) ;
		\draw ($(c0)+(90-\angTheta/2:\lenR*2/3)$) node {$\vartheta$};
		\end{tikzpicture}
	}
%%%%%%%%%%%%%%%%%%%%%%%%%%%%%%%%%%%%%%%%%%		
	\caption{\label{fig:schematic0} (Color online) Left: Sketch of the geometry, not to scale. A Janus particle of Radius $R$ with different reflection properties on its two hemispheres is suspended in a gas between two parallel plates of temperatures $T_l$ and $T_h$, separated by a distance $h$ with $h\gg 2R$. The orientation of the Janus particle is characterized by the unit vector $\mathbf{n}_p$ normal to the equatorial plane separating the two hemispheres of the particle. The orientation of $\mathbf{n}_p$ with respect to a Cartesian coordinate system with $z$-axis normal to the bounding plates can be given as polar and azimuthal angles $\theta$ and $\phi$, with $\mathbf{n}_p = \sin\theta\cos\phi\,\mathbf{e}_x + \sin\theta\sin\phi\,\mathbf{e}_y + \cos\theta\,\mathbf{e}_z$. 
		\\
		Right: Points $\mathbf{r}=\mathbf{r}(\vartheta,\varphi)$ on the surface of the particle are parametrized by the polar and azimuthal angles $\vartheta$ and $\varphi$ with $\mathbf{r}(\vartheta,\varphi) = R(\sin\vartheta\cos\varphi\,\mathbf{e}_x + \sin\vartheta\sin\varphi\,\mathbf{e}_y + \cos\vartheta\,\mathbf{e}_z)$. This defines the usual local coordinate system with unit vectors $\mathbf{e}_r=\mathbf{r}/R$, $\mathbf{e}_\vartheta = \partial_\vartheta \mathbf{r}(\vartheta,\varphi)/R$ and $\mathbf{e}_\varphi=\mathbf{e}_r\times\mathbf{e}_\vartheta$. The surface normal $\mathbf{n}$ coincides with $\mathbf{e}_r(\vartheta, \varphi)$.}
\end{figure}
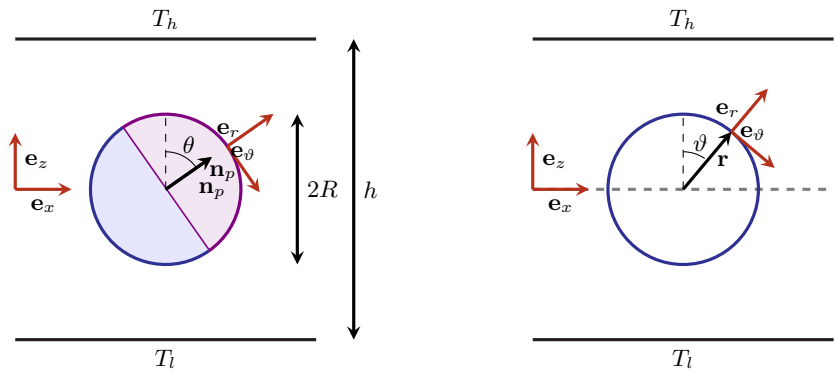

Thermophoresis in gases, i.e. the motion of aerosol particles along a gradient in temperature, has been applied both for preventing and inducing particle deposition on heated or cooled surfaces from the gas phase, as already suggested by \citet{Aitken_1884} in his seminal study of the phenomenon. In the present paper we investigate whether thermophoretic deposition can be used for decorating surfaces with non-symmetric nanoparticles oriented with respect to the applied temperature gradient. As a model system we consider a spherical Janus particle \cite{deGennes_1992}, figure \ref{fig:schematic0}, where each hemisphere is characterised by a different accommodation coefficient for the reflection of gas molecules (giving it a two-faced appearance similar to depictions of the god Janus). In the limit of large Knudsen numbers, $\mathrm{Kn}=\ell/(2R)\gg1$, when the mean free path $\ell$ of gas molecules is much larger than the particle diameter $2R$, the force and torque on the particle can be calculated analytically. This allows identifying the interplay between thermophoretic motion and particle alignment. Without translation the Janus particle is oriented with its more diffuse side pointing preferentially in the direction of the lower temperature. However, a translation of the Janus particle through the surrounding gas results in a torque aiming to orient the particles' more diffuse side in direction opposite to the direction of motion. Since the thermophoretic force leads to a net motion towards the colder gas, this thus leads to a weakening of the particle alignment.

The earliest estimate for the thermophoretic force on a particle at large Knudsen numbers seems to be due to \citet{Einstein_1924}. More exact calculations for the force and drag on a homogeneous sphere at large Knudsen numbers were later performed by \citet{Waldmann_1959} and simultaneously by \citet{Bakanov_1960}, allowing the determination of the thermophoretic velocity in this limit. Our analytical calculations largely follow these early presentations. Notable extensions of these results to rotating particles of various shape were presented by Borg, S\"oderholm and Ess{\'e}n, \cite{Borg_2001, Borg_2003}. Extensive reviews of analytical, numerical and experimental results on the thermophoretic motion of particles in the gas phase, including the transition flow and slip-flow regime, can be found in \cite{Davis_2002, Zheng_2002, Sone_2007, Young_2011}. 

In the past few years it has become feasible to synthesize micro- or nanoparticles showing large deviations from spherical symmetry, either by their geometry or by their surface properties \cite{Glotzer_2007, Lorke_2012}. In that context Janus particles, composed of two hemispheres with different properties, have probably received the most attention \cite{Hu_2012, Walther_2013, Zhang_2017}. In the context of the present work, only the surface properties of the two hemispheres are relevant. Two hemispheres with different momentum accommodation coefficients can be formed by varying the surface roughness \cite{Agrawal_2008, Cao_2009} or by covering a part of the surface with a thin film \cite{Seo_2013, Seo_2014}. While the motion of Janus particles in liquids have been widely discussed \cite{Jiang_2010, Bickel_2013}, little attention was given to the theoretical description of their transport in the gas phase. Notable exceptions are the studies by \citet{Rohatschek_1984} and \citet{Beresnev_2012} considering the photophoretic force due to inhomogeneous heating by absorption of electromagnetic radiation on a Janus particle with different accomodation coefficients on its hemispheres in the limit of large Knudsen numbers. We complement these results to arbitrarily translating and rotating Janus particles in a thermal gradient.

This paper is organised as follows: In section \ref{sec:transfer} we determine expressions for the force and torque on a Janus particle. Section \ref{sec:ChapmanEnskog} considers a Janus particle in a gas described by a Chapman-Enskog distribution, and the particle motion is illustrated by numerical solutions of the Langevin equation, highlighting the interplay between thermophoretic motion and orientation in the temperature field. In section \ref{sec:transferKnInf} a Janus particle in the collisionless regime between two parallel surfaces is considered analytically. These results are complemented using the Direct Simulation Monte Carlo (DSMC) method in section \ref{sec:DSMC}, where also finite Knudsen numbers are considered. 

%%%%%%%%%%%%%%%%%%%%%%%%%%%%%%% figure %%%%%%%%%%%%%%%%%%%%%%%%%%%%%%%
%%%%%%%%%%%%%%%%%%%%%%%%%%%%%%%%%%%%%%%%%%%%%%%%%%%%%%%%%%%%%%%%%%%

%%%%%%%%%%%%%%%%%%%%%%%%%%%%%%%%%%%%%%%%%%%%%%%%%%%%%%%%
\section{Momentum transfer between gas and particle}\label{sec:transfer}

In order to calculate the force on the Janus particle, we assume that the mean free path of gas molecules is much larger than the diameter of the particle, such that the phase space distribution function $f(\mathbf{r}, \mathbf{c})$ of gas molecules impinging onto the particle can be regarded as unchanged by the presence of the particle. We will also assume that the gas only contains one kind of molecules having mass $m$ which are much lighter than the Janus particle, such that the particle's recoil during a collision with a molecule results in an infinitesimal change in its velocity. For the interaction of the gas molecules with the particle surface we assume a Maxwell-type boundary condition with a position dependent accommodation coefficient $a(\mathbf{r})$, which can be interpreted as the fraction $a$ of the incoming molecules that is reflected diffusely, while the fraction $(1-a)$ is reflected specularly. In particular, for a surface moving at velocity $\mathbf{u}_w$ and with wall temperature $T_p$, the phase space distribution of the outgoing molecules obeys the boundary condition
\begin{equation}
f(\mathbf{r}, \mathbf{c}) = a(\mathbf{r})f^\mathrm{(d)}(\mathbf{r}, \mathbf{c}; \mathbf{u}_w)
	+(1-a(\mathbf{r})) f(\mathbf{r},\mathbf{c}-2\mathbf{n}((\mathbf{c}-\mathbf{u}_w)\cdot\mathbf{n})),
	\quad \text{for} \quad  (\mathbf{c}-\mathbf{u}_w)\cdot\mathbf{n} > 0
\end{equation}
where $\mathbf{n}$ is the outward unit normal vector at the surface pointing into the gas phase. Diffusely reflected molecules are characterised by a Maxwellian distribution
\begin{equation}
f^\mathrm{(d)}(\mathbf{r}, \mathbf{c}; \mathbf{u}_w) = \nu(\mathbf{r}) \frac{2\beta_p^2}{\pi} e^{-\beta_p(\mathbf{c}-\mathbf{u}_w)^2},
\end{equation}
where we have used the abbreviation $\beta_p=m/(2k T_p)$, with the Boltzmann constant $k$, and where
\begin{equation}
\nu(\mathbf{r}) = - \int_{(\mathbf{c}-\mathbf{u}_w)\cdot\mathbf{n}<0} ((\mathbf{c}-\mathbf{u}_w)\cdot\mathbf{n}) f(\mathbf{r},\mathbf{c}) d^3c
\end{equation}
is the molecule flux density at position $\mathbf{r}$ on the surface, i.e. the number of gas molecules impinging on the surface per unit area and per unit time. The force density on the wall due to the interaction with gas molecules is the momentum flux per unit area
\begin{equation}
\mathbf{\Pi}(\mathbf{r}) = - \int ((\mathbf{c}-\mathbf{u}_w)\cdot\mathbf{n}) m\mathbf{c} f(\mathbf{r},\mathbf{c}) d^3c
\end{equation}
carried by molecules towards and away from the wall.

The integrals are conveniently evaluated in an inertial frame of reference co-moving with the momentary velocity $\mathbf{u}$ of the particle and with the origin of the spatial coordinate system in the center of the sphere. Thus, when the particle spins at an angular velocity $\boldsymbol{\omega}$, the wall velocity at position $\mathbf{r}$ on the surface, $\mathbf{u}_w=\boldsymbol{\omega}\times\mathbf{r}$, will be orthogonal to the outward normal, $\mathbf{n}\equiv\mathbf{e}_r=\mathbf{r}/R$, at the surface, $\mathbf{n}\cdot\mathbf{u}_w=0$. The force density at the surface can then be written as 
\begin{equation} \label{eq:momFlux}
\mathbf{\Pi}(\mathbf{r}) = a(\mathbf{r})\left( \mathbf{\Pi}_\mathrm{in}(\mathbf{r}) + \mathbf{\Pi}_\mathrm{out}^\mathrm{(d)}(\mathbf{r}) \right) + (1-a(\mathbf{r}))\; 2(\mathbf{\Pi}_\mathrm{in}(\mathbf{r})\cdot\mathbf{n}) \mathbf{n},
\end{equation} 
with
\begin{align}
\mathbf{\Pi}_\mathrm{in}(\mathbf{r}) &= -\int_{\mathbf{c}\cdot\mathbf{n}<0} (\mathbf{c}\cdot\mathbf{n}) (m\mathbf{c}) f(\mathbf{r}, \mathbf{c}) d^3\mathbf{c}, \label{eq:momFluxIn}\\
\mathbf{\Pi}_\mathrm{out}^\mathrm{(d)}(\mathbf{r}) & = -\int_{\mathbf{c}\cdot\mathbf{n}>0} (\mathbf{c}\cdot\mathbf{n}) (m\mathbf{c}) f^\mathrm{(d)}(\mathbf{r}, \mathbf{c}; \boldsymbol{\omega}\times\mathbf{r} ) d^3\mathbf{c} \label{eq:momFluxOut}\\
& = -\int_{\mathbf{c}\cdot\mathbf{n}>0} (\mathbf{c}\cdot\mathbf{n}) (m(\mathbf{c}+\boldsymbol{\omega}\times\mathbf{r}))  f^\mathrm{(d)}(\mathbf{r}, \mathbf{c}; \mathbf{0}) d^3\mathbf{c}\\
& = - \nu(\mathbf{r}) m \left[ \frac{1}{2}\sqrt{\frac{\pi}{\beta_p}}\mathbf{n} + \boldsymbol{\omega}\times\mathbf{r}\right],\\ \label{eq:partFlux}
\nu(\mathbf{r}) &= - \int_{\mathbf{c}\cdot\mathbf{n}<0} (\mathbf{c}\cdot\mathbf{n}) f(\mathbf{r},\mathbf{c}) d^3c.
\end{align}
Here, $\mathbf{\Pi}_\mathrm{in}$ is the inward momentum flux at the surface, with its normal component contributing with a factor 2 to the total momentum transferred on specular surfaces. Similarly, $\mathbf{\Pi}_\mathrm{out}^\mathrm{(d)}$ is readily identified as the outward momentum flux on a diffuse surface where, due to rotation of the particle, on average one molecule carries away the additional momentum $m\boldsymbol{\omega}\times\mathbf{r}$ tangential to the surface compared to the instantaneous rest frame of the surface element.

The force and torque on the particle are obtained by integrating $\mathbf{\Pi}(\mathbf{r})$ and $\mathbf{r}\times\mathbf{\Pi}(\mathbf{r})$ over the surface of the particle, respectively. For this, the phase space distribution function $f(\mathbf{r}, \mathbf{c})$ of the molecules impinging on the particle surface has to be specified. In the following sections we will consider two cases: (i) a Chapman-Enskog distribution with a thermal gradient and (ii) a distribution between diffusely reflecting parallel plates held at different temperatures in the collisionless regime. 

Figure \ref{fig:schematic0} illustrates the geometry and the coordinate system used. For both cases described above, we assume a Janus particle with radius $R$ located between two parallel plates at distance $h$. For our analytical calculations, the coordinate system is chosen such that its origin lies at the momentary center of the sphere. The $z$-axis of this coordinate system lies in direction normal to the plates, while the $x$- and $y$-axes span a plane parallel to them. Points $\mathbf{r}=\mathbf{r}(\vartheta,\varphi) = R(\sin\vartheta\cos\varphi\,\mathbf{e}_x + \sin\vartheta\sin\varphi\,\mathbf{e}_y + \cos\vartheta\,\mathbf{e}_z)$ on the surface of the particle are parametrized by the polar and azimuthal angles $\vartheta$ and $\varphi$ with $0\le\vartheta\le\pi$ and $0\le\varphi<2\pi$. This defines the usual local coordinate system with unit vectors $\mathbf{e}_r=\mathbf{r}/R$, $\mathbf{e}_\vartheta = \partial_\vartheta \mathbf{r}(\vartheta,\varphi)/R$ and $\mathbf{e}_\varphi=\mathbf{e}_r\times\mathbf{e}_\vartheta$ on the surface of the sphere. The surface normal $\mathbf{n}$ coincides with $\mathbf{e}_r(\vartheta, \varphi)$. The orientation of the Janus particle is characterized by the unit vector $\mathbf{n}_p$ normal to the equatorial plane separating the two hemispheres of the particle. The momentum accommodation coefficient $a(\mathbf{r})$ is $a^+$ on the upper hemisphere with $\mathbf{r}\cdot\mathbf{n}_p >0$ and $a^-$ on the lower hemisphere. The orientation of $\mathbf{n}_p$ with respect to the Cartesian coordinate system is parametrized by the polar and azimuthal angles $\theta$ and $\phi$ with $\mathbf{n}_p=\mathbf{n}_p(\theta,\phi) = \sin\theta\cos\phi\,\mathbf{e}_x + \sin\theta\sin\phi\,\mathbf{e}_y + \cos\theta\,\mathbf{e}_z$. 

\section{Chapman-Enskog-Distribution, $\ell \ll h$}\label{sec:ChapmanEnskog}

\subsection{Force and torque on a Janus particle}\label{sec:PhaseSpace_ChapmanEnskog}

In this section we focus on the situation that the plate distance $h$ is much larger than the mean free path of the gas molecules, and the Janus particle is located far away from both plates. Under these circumstances the state of the gas at rest is characterised by the molecule density, $n$, temperature, $T$, pressure, $p=nkT$, and conductive heat flux $\mathbf{q}=-\lambda \boldsymbol{\nabla}T$, where $\lambda$ is the thermal conductivity of the gas. Let $\tilde{f}(\mathbf{r},\tilde{\mathbf{c}})$ be the phase space density in the rest frame of the gas, where for the moment we have denoted quantities in this frame of reference by a tilde. To first order in the temperature gradient it can be obtained from the Boltzmann equation via a Chapman-Enskog expansion as \cite{Landau_1983_X, Davis_2002, Waldmann_1959, Bird_1994} 
\begin{align}
\tilde{f}(\mathbf{r},\tilde{\mathbf{c}}) &=n\left(\frac{\beta}{\pi}\right)^{3/2} e^{-\beta \tilde{c}^2} \left[1-\frac{2}{5}\frac{m\lambda}{n(kT)^2}(\beta \tilde{c}^2-\tfrac{5}{2})\,\tilde{\mathbf{c}}\cdot\boldsymbol{\nabla} T\right],
\end{align}
with the abbreviation $\beta = m/(2kT)$. The relation between the molecular velocities $\tilde{\mathbf{c}}$ in the rest frame of the gas and their velocities $\mathbf{c}$ in the inertial frame moving with the momentary velocity $\mathbf{u}$ of the particle relative to the gas is given by a Galilean transformation, $\mathbf{c}=\tilde{\mathbf{c}}-\mathbf{u}$, and thus the distribution function in the rest frame of the particle is $f(\mathbf{r},\mathbf{c}) = \tilde{f}(\mathbf{r},\tilde{\mathbf{c}}) = \tilde{f}(\mathbf{r},\mathbf{c}+\mathbf{u})$. For our purposes it suffices to expand this to first order in the particle velocity and only keep the lowest order terms in $\boldsymbol{\nabla} T$ and $\mathbf{u}$
\begin{align}\label{eq:ChapmanEnskogDistribution}
f(\mathbf{r},\mathbf{c}) &\approx n\left(\frac{\beta}{\pi}\right)^{3/2} e^{-\beta c^2} \left[1-\frac{2}{5}\frac{m\lambda}{n(kT)^2}(\beta c^2-\tfrac{5}{2})\,\mathbf{c}\cdot\boldsymbol{\nabla} T - 2\beta\,\mathbf{c}\cdot\mathbf{u}\right].
\end{align}
With this velocity distribution function the molecule flux density on the surface of the sphere is calculated using equation (\ref{eq:partFlux}),
\begin{align}\label{eq:nu_ChapmanEnskog}
\nu(\mathbf{r}) = \frac{n}{2} \left[ \frac{1}{\sqrt{\pi\beta}} + \mathbf{n}\cdot\mathbf{u} \right],
\end{align}
and the inward momentum flux, eq. (\ref{eq:momFluxIn}), becomes
\begin{align}
\mathbf{\Pi}_\mathrm{in}(\mathbf{r}) 
&= - \frac{nm}{2\beta}\left[ \frac{1}{2} \mathbf{n} + \sqrt{\frac{\beta}{\pi}} 
\left( \mathbf{I} + \mathbf{n}\otimes\mathbf{n} \right) \cdot \hat{\mathbf{u}} \right],
\end{align}
where we have used the abbreviation 
\begin{align}
\hat{\mathbf{u}} = \mathbf{u} + \left[ \lambda/(5p) \right] \boldsymbol{\nabla} T = \mathbf{u} - \mathbf{q}/(5p).
\end{align}
The tensor product obeys $\mathbf{a}\otimes\mathbf{b} \cdot \mathbf{c} = \mathbf{a} (\mathbf{b} \cdot \mathbf{c})$ and $\mathbf{I}$ is the identity matrix with $\mathbf{I}\cdot\mathbf{a}=\mathbf{a}\cdot\mathbf{I}=\mathbf{a}$. Collecting terms we can write for the differential force per surface element (with $a=a(\mathbf{r})$)
\begin{align}\label{eq:dFmu}
\begin{split}
\mathbf{\Pi}(\mathbf{r}) =& -p\left[ \tfrac{1}{2}\left(2-a+a\sqrt{T_p/T}\right)\, \mathbf{n} + \sqrt{\frac{\beta}{\pi}} \left\{ a \hat{\mathbf{u}} + (4-3a) \mathbf{n}\otimes\mathbf{n}\cdot \hat{\mathbf{u}} + \sqrt{(T_p/T)}\frac{\pi}{2} a \, \mathbf{n}\otimes\mathbf{n}\cdot \mathbf{u} \right\}  \right] \\
&- pa\left( \sqrt{\frac{\beta}{\pi}} + \beta \, \mathbf{n}\cdot\mathbf{u} \right) \, \boldsymbol{\omega}\times(R\mathbf{n}).
\end{split}
\end{align}
The net force and torque are calculated by integrating $\mathbf{\Pi}(\mathbf{r})$ and $\mathbf{r}\times\mathbf{\Pi}(\mathbf{r})$ over the surface of the sphere. This requires the integration of the tensors products of the surface-normals up to third order over each hemisphere, presented in appendix \ref{sec:Integrals_1}. With these integrals we obtain for the force on the Janus sphere ($S_0=4\pi R^2$),
\begin{align}\label{eq:F_ChapmanEnskog}
\begin{split}
\mathbf{F} =& - \frac{4}{3} S_0 p \sqrt{\frac{\beta}{\pi}} \left[\hat{\mathbf{u}} + \frac{\pi}{8}\sqrt{\frac{T_p}{T}}\,\tfrac{1}{2}(a^+ + a^-) \mathbf{u} \right] \\
&+\frac{1}{8} S_0 p \, (a^+ - a^-) \left(1-\sqrt{\frac{T_p}{T}} \right) \mathbf{n}_p \\
&-\frac{1}{4} S_0 p \sqrt{\frac{\beta}{\pi}}(a^+ - a^-) \, (R\boldsymbol{\omega}\times\mathbf{n}_p) \\
&-\frac{1}{3} S_0 p \, \beta \,\tfrac{1}{2}(a^+ + a^-) \, (R\boldsymbol{\omega}\times\mathbf{u} ).
\end{split}
\end{align}
The first line in this expression agrees with the results of Bakanov and Derjaguin \cite{Bakanov_1960} and Waldmann \cite{Waldmann_1959} for the thermophoretic and drag forces on a sphere with surface-temperature $T_P=T$ and homogeneous accommodation coefficient $a=a^+=a^-$. Under these conditions the net force on the particle vanishes when it moves with a drift velocity $\mathbf{u}_d=\mathbf{q}/[5p(1+a\pi/8)]$. The second line is purely due to diffuse reflection from a sphere with a temperature different from its surroundings. The third line is due to the inhomogeneous rotational drag on the diffuse surface, while the fourth line corresponds to a 'negative' Magnus force due to impinging particles being reflected predominantly in direction of rotation from the diffuse surface (\cite{Wang_1972}, \cite{Ivanov_1980}, \cite{Borg_2003}, \cite{Weidman_2004}). Note that in these expressions $S_0p$ serves as the natural scale for the force, while the rotational and translational velocities, $\mathbf{u}$ and $R\boldsymbol{\omega}$, are scaled with the characteristic thermal velocity $\bar{c}=1/\sqrt{\beta}=\sqrt{2kT/m}$ of the gas molecules.

Similarly, integrating the torque density, $d\mathbf{M}(\mathbf{r})=\mathbf{r}\times\mathbf{\Pi}(\mathbf{r})dS$ with $\mathbf{r}=R \mathbf{n}$, over the surface of the sphere yields
\begin{align}\label{eq:M_ChapmanEnskog}
\begin{split}
\mathbf{M} =& -\frac{2}{3} S_0 p R^2 \sqrt{\frac{\beta}{\pi}} \tfrac{1}{2}(a^+ + a^-) \boldsymbol{\omega} \\
&- \frac{1}{4} S_0 p R \sqrt{\frac{\beta}{\pi}}  (a^+ - a^-) (\mathbf{n}_p\times\hat{\mathbf{u}}) \\
&+ \frac{1}{16} S_0 p R^2 \beta \, (a^+ - a^-)\big[ -3 (\mathbf{u}\cdot\mathbf{n}_p) \boldsymbol{\omega} + \mathbf{u} (\boldsymbol{\omega}\cdot\mathbf{n}_p) + (\mathbf{u}\cdot\boldsymbol{\omega}-(\mathbf{u}\cdot\mathbf{n}_p)(\boldsymbol{\omega}\cdot\mathbf{n}_p))\mathbf{n}_p \big].
\end{split}
\end{align}
In this expression the first line corresponds to rotational drag, and for a sphere with a homogeneous accommodation coefficient $a=a^+ = a^-$ we recover $\mathbf{M}=-\alpha_\omega \boldsymbol{\omega}$, with the familiar friction coefficient $\alpha_\omega = \frac{2\pi}{3}aR^4 n\sqrt{8mkT/\pi}$, \cite{Loyalka_1992}.

Note that on fully specular surfaces the net momentum transfer is purely normal to the surface, since the normal component of velocity is reversed while the tangential component is conserved upon reflection. On diffusely reflecting surfaces the outward momentum flux consists of a purely normal part due to the diffuse reflection and a purely tangential part due to the rotation of the sphere. Since the torque is the surface integral of $d\mathbf{M}(\mathbf{r})=\mathbf{r}\times\mathbf{\Pi}(\mathbf{r})dS$, it is seen that only diffusely reflecting surfaces contribute to the torque and in particular that the torque is independent of the temperature distribution on the surface of the sphere even for non-uniform temperature distributions.

\subsection{Thermophoresis and  orientation of a Janus-sphere\label{sec:thermophoresis_ChapmanEnskog}}

The motion of the Janus particle is dictated by the bombardment of gas molecules resulting in an average force and torque as calculated above. Due to the stochastic nature of the bombardment, the momentary force and torque on the particle are subject to fluctuations, leading to center-of-mass and orientational diffusion. In the following, we will choose the spherical Janus particle to have a homogeneous density $\rho_p$, such that its mass and moment of inertia are $m_p=(4\pi/3)\rho_p R^3$ and $I_p =(2/5)m_pR^2$, respectively. From the equipartition theorem, assuming the particle to be almost in thermal equilibrium with the surrounding gas, it will have a typical thermal translational and rotational velocity of $\sqrt{2kT/m_P}$ and $R\omega\sim R\sqrt{2kT/I_P}$, respectively, where $T$ is the gas temperature at the particle position. The characteristic velocity of gas molecules, $\bar{c}=1/\sqrt{\beta}=\sqrt{2kT/m}$, is much larger than both these particle-velocities, since by assumption the particle mass is much larger than the molecular mass. The last term in each of the expressions for $\mathbf{F}$ and $\mathbf{M}$ are thus expected to be negligible compared to the other terms, since they are of second order in the ratio of particle and molecular velocities, while the others are of first order in these ratios. This also remains true for a particle moving at its thermophoretic drift velocity $\mathbf{u}_d \simeq \mathbf{q}/(5p)$. Note that for the force this corresponds to the Magnus force on the particle being negligible for the particle moving at its thermal velocity. 

For a particle moving much more slowly than the typical molecular velocities in the gas it is thus admissible to neglect the terms quadratic in $\boldsymbol{\omega}$ and $\mathbf{u}$. When further assuming the temperature of the particle to be the same as that of the surrounding gas phase, $T_p=T$, the force and torque on the sphere have the form
\begin{align}
\mathbf{F} &= - \alpha_u \mathbf{u} + A_q \mathbf{n}_q + A_\omega (\mathbf{n}_p \times \boldsymbol{\omega}) \\
\mathbf{M} &=-\alpha_\omega \boldsymbol{\omega} + B_q (\mathbf{n}_p\times \mathbf{n}_q) + B_u (\mathbf{n}_p\times\mathbf{u}) \label{eq:M_ChapmanEnskog_Parametric}
\end{align}
where $\mathbf{n}_q= \boldsymbol{q}/|\boldsymbol{q}|$ is the unit vector pointing in direction of the diffusive heat flux, and the coefficients $\alpha_\omega$, $A_q$, $A_\omega$, $\alpha_\omega$, $B_q$ and $B_u$ can be read off from equations (\ref{eq:F_ChapmanEnskog}) and (\ref{eq:M_ChapmanEnskog}). When we allow for fluctuations in the force and torque, we obtain the Newton-Euler-Langevin equations \cite{McConnell_1980, Coffey_2012} dictating the particle motion
\begin{align}
m_p d\mathbf{u} &= - \alpha_u \mathbf{u}\,dt + A_q \mathbf{n}_q\,dt + A_\omega (\mathbf{n}_p \times \boldsymbol{\omega})\,dt +\sqrt{2\alpha_u kT} d\mathbf{W}_u \label{eq:LangevinU}\\
I_p d\boldsymbol{\omega} &=-\alpha_\omega \boldsymbol{\omega}\,dt + B_q (\mathbf{n}_p\times \mathbf{n}_q)\,dt + B_u (\mathbf{n}_p\times\mathbf{u})\,dt +\sqrt{2\alpha_\omega kT} d\mathbf{W}_\omega \label{eq:LangevinOmega}\\
d\mathbf{n}_p &=\boldsymbol{\omega}\times\mathbf{n}_p\,dt \label{eq:LangevinN}
\end{align}
where the fluctuations are assumed to be Gaussian white noise with \cite{Jacobs_2010, Gardiner_2004, Kampen_2007}
\begin{align}
\langle d\mathbf{W}_\sigma (t) \otimes d\mathbf{W}_{\sigma'} (t') \rangle = \mathbf{I} \delta(t-t') \delta_{\sigma\sigma'} dt,\qquad \sigma,\sigma'\in\{u,\omega\}.
\end{align}
The set of stochastic differential equations can be discretized via the Euler algorithm such that for a finite timestep $\Delta t$ the incremental fluctuations are $\Delta\mathbf{W}_{\sigma}=\sqrt{\Delta t}\,\boldsymbol{{\cal N}}_\sigma(0,1)$, where $\boldsymbol{{\cal N}}_\sigma (0,1)$ are vectors containing independent normally distributed random variables with mean 0 and variance 1 at each timestep. In appendix \ref{sec:nondimEuler} equations (\ref{eq:LangevinU})-(\ref{eq:LangevinN}) are stated in non-dimensional form adequate for a numerical analysis.

For a particle at rest, $\mathbf{u}=0$, the Euler equation describing the orientation of the particle are equivalent to those of a damped magnetic dipole rotating in an applied magnetic or electric field \cite{McConnell_1980, Coffey_2012}, where the interaction energy of the dipole and the field can be described as a potential
\begin{equation}\label{eq:potential}
V_q(\mathbf{n}_p) = -B_q (\mathbf{n}_p \cdot \mathbf{n}_{q} ) = -B_q \cos\tilde{\theta},\qquad B_q = \frac{S_0R}{20}\sqrt{\frac{\beta}{\pi}}(a^+-a^-)|\mathbf{q}|,
\end{equation}
and where $\tilde{\theta}$ is the angle between the particle axis and the heat flux. In equilibrium, the angular probability distribution function $p(\tilde{\theta},\tilde{\phi})$ for the orientation of the particle in polar coordinates is thus given by the Boltzmann distribution
\begin{equation}\label{eq:BoltzmannThetaPhi}
p(\tilde{\theta}, \tilde{\phi}) = {\cal N} \exp[-B_q \cos(\tilde{\theta})/(kT)], 
\qquad {\cal N}^{-1}= \int_0^\pi \int_0^{2\pi} p(\tilde{\theta},\tilde{\phi})\sin\tilde{\theta}\;d\tilde{\phi}\;d\tilde{\theta} = 4\pi \frac{\sinh(B_q/(kT))}{B_q/(kT)}
\end{equation}
and the marginal distribution, $p(\tilde{\theta})=\int_0^{2\pi} p(\tilde{\theta},\tilde{\phi})\sin\tilde{\theta}\;d\tilde{\phi}$, for the polar angle $\tilde{\theta}$ is
\begin{equation}\label{eq:BoltzmannTheta}
p(\tilde{\theta}) = \tilde{\cal N} \sin\tilde{\theta} \exp[-B_q \cos(\tilde{\theta})/(kT)],\qquad \tilde{{\cal N}}^{-1}= \int_0^\pi p(\tilde{\theta})\;d\tilde{\theta} = 2 \frac{\sinh(B_q/(kT))}{B_q/(kT)}.
\end{equation}
When the 'northern' hemisphere is more diffusely reflecting than the southern hemisphere, $a^+>a^-$, i.e. $B_q>0$,  the Janus-particle thus aligns with the heat flux, i.e. with the more diffuse side pointing towards the colder region. Note that by the same reasoning, and since $B_q$ and $B_u$ have the opposite sign, a particle moving at a constant velocity with respect to an isothermal gas will orient preferably with the more diffuse side opposite to the direction of motion.

We next turn to the translational motion of the particle, governed by equation (\ref{eq:LangevinU}). At steady state, the average force $\langle \mathbf{F} \rangle$ on the particle vanishes. The next-to-last term in equation (\ref{eq:LangevinU}) describes the contribution of the coupling between particle rotation and orientation to the force. It is plausible that for any orientation $\boldsymbol{n}_p$ the rotational velocity has no preferred direction, i.e. $\langle \boldsymbol{\omega} \rangle|_{\boldsymbol{n}_p} = 0$. Thus, when averaging equation (\ref{eq:LangevinU}) over the entire phase space, the contribution of this term vanishes, and the drift velocity becomes
\begin{align}\label{eq:uDriftChapmanEnskog}
\mathbf{u}_d = \langle \mathbf{u} \rangle = \frac{A_q}{\alpha_u}\mathbf{n}_{q} = \frac{\mathbf{q}}{5p} \left[ 1 + \frac{\pi}{16}\left(a^++a^-\right)\right]^{-1}.
\end{align}

To find the coupling of the drift velocity to the rotation, we average the torque $\mathbf{M}$ over velocity space. Since the drift velocity is colinear with the heat flux, the last two terms in equation (\ref{eq:M_ChapmanEnskog_Parametric}) can be combined to an effective torque such that
\begin{align}
\mathbf{M}^{(\text{eff})} = -\alpha_\omega \boldsymbol{\omega} + B_q^{(\text{eff})} (\mathbf{n}_p\times \mathbf{n}_{q}),
\end{align}
where we have introduced the effective coupling strength
\begin{align}\label{eq:potEffCE}
B_q^{(\text{eff})} = \left(B_q + B_u \frac{A_q}{\alpha_u} \right) = B_q \left(1-\left[ 1 + \frac{\pi}{16}\left(a^++a^-\right)\right]^{-1}\right).
\end{align}
The translation of the particle thus weakens the torque aligning the particle with the heat flux. This is expected, since the drift is also in direction of the heat flux, such that the torque due to the heat flux and due to the drift velocity partially compensate. The situation can be interpreted as a dipole in a potential as in equation (\ref{eq:potential}), where the coupling strength $B_q$ has been replaced by the effective coupling strength $B_q^{(\text{eff})}$, and the angular distribution of the particle is again dictated by a corresponding Boltzmann distribution as in (\ref{eq:BoltzmannTheta}). For the case of maximal dissimilarity between the hemispheres of the Janus particle, $a^+=1$ and $a^-=0$, we obtain $B_q^{(\text{eff})}=0.164\,B_q$.

\begin{figure}
\includegraphics[width=7cm]{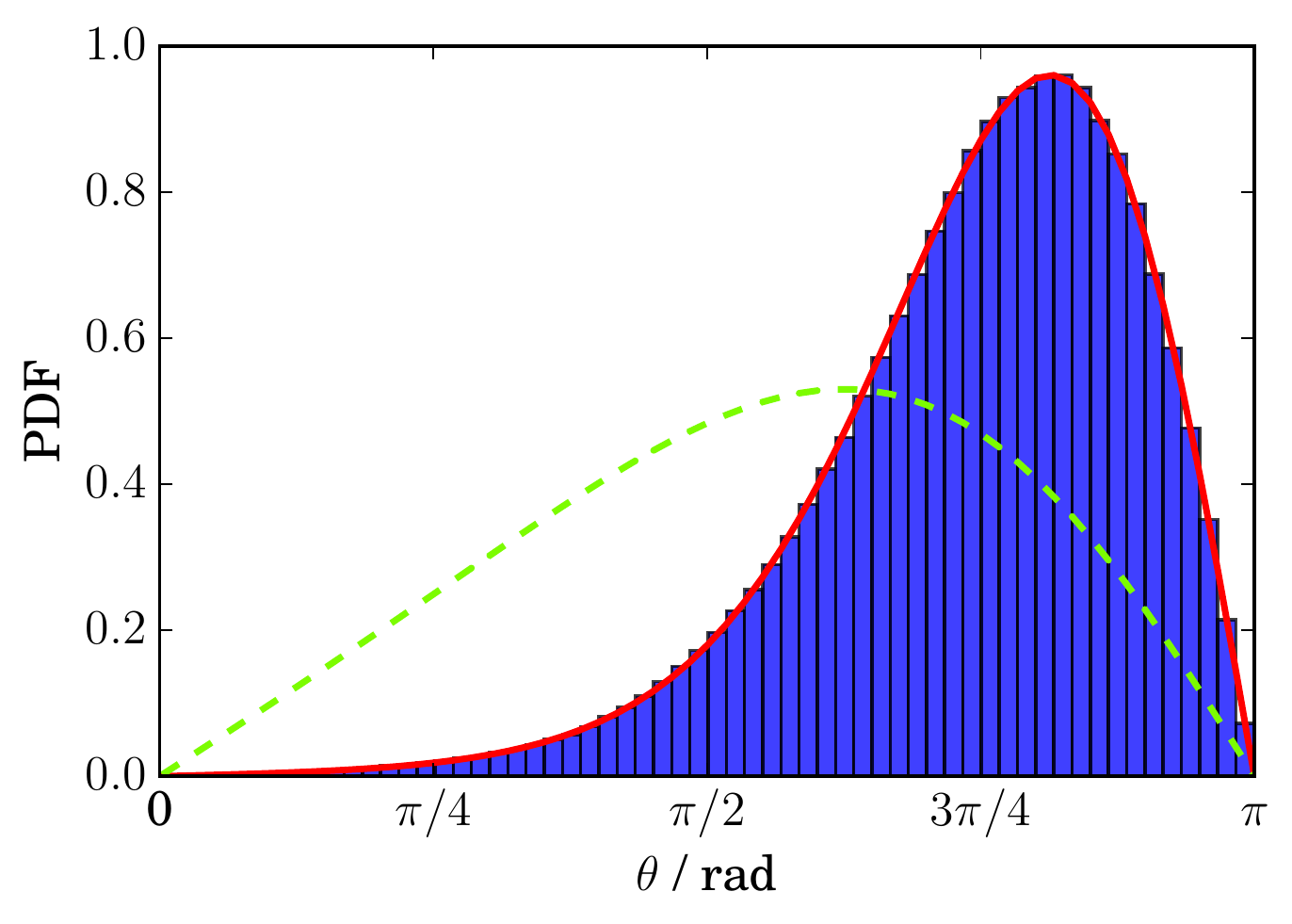} \hspace{0.5cm}
\includegraphics[width=7cm]{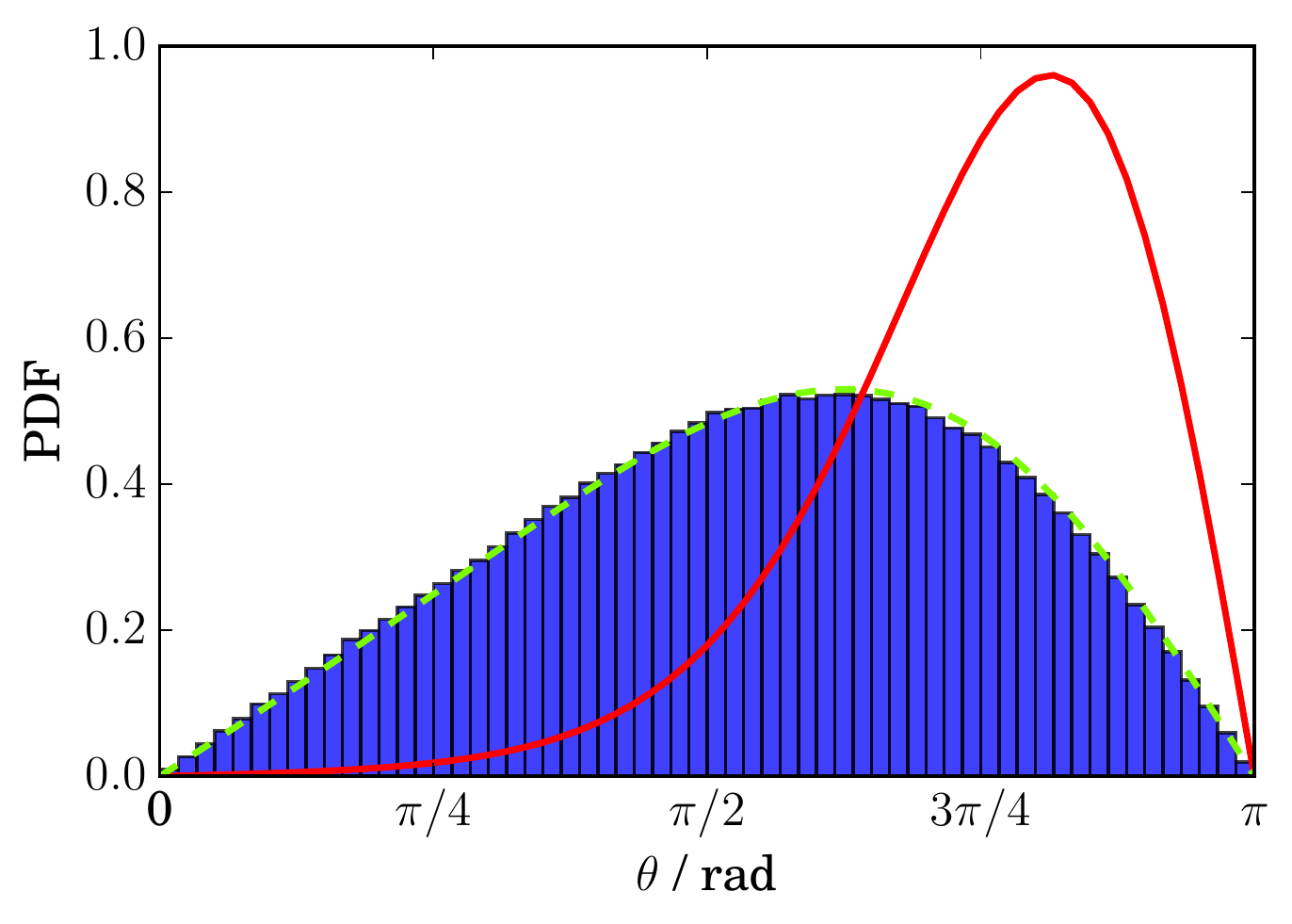}
\caption{\label{fig:Langevin} (Color online) Probability density functions (PDF) of the Janus sphere orientation ($\cos\theta=\mathbf{n}_p\cdot\mathbf{e}_z$). Left: stationary sphere, right: translating sphere. Blue histograms: numerical results from the Newton-Euler-Langevin equation; Red lines: prediction from the Boltzmann distribution for $\mathbf{u}=0$; Green dashes: prediction from the Boltzmann distribution for $\mathbf{u}=\mathbf{u}_d$. $T=300$~K, $R=500$~nm, $\boldsymbol{\nabla} T=5$~K/mm~$\mathbf{e}_z$, $\rho_p = 1000$~kg/m$^3$, $a^+=1$, $a^-=0$, $\rho_g=0.011$~kg/m$^3$, $m=6.63\cdot10^{-26}$~kg, $\lambda = 18$~mW/(mK), $\mathbf{q}=-\lambda \boldsymbol{\nabla}T$.
}
\end{figure}

In order to validate the qualitative reasoning leading to equations (\ref{eq:uDriftChapmanEnskog}-\ref{eq:potEffCE}), we discretize the Newton-Euler-Langevin equations (\ref{eq:LangevinU}-\ref{eq:LangevinN}) using an Euler algorithm (see appendix \ref{sec:nondimEuler} for details) and simulate the motion of the sphere. The equations are non-dimensionalised using a timescale $\tau=m_p/(S_0p\sqrt{\beta/\pi})$, a velocity scale $U_0=R/\tau$ and a frequency scale $\Omega_0=1/\tau$, such that $\mathbf{u}/U_0$ and $\boldsymbol{\omega}/\Omega_0$ become the dimensionless velocity and angular frequency. As an example, we consider a sphere of radius $R=500$ nm and density $\rho_p = 1000$ kg/m$^3$, with a diffusely reflecting northern hemisphere and a specularly reflecting southern hemisphere, $a^+=1$, $a^-=0$. The sphere is surrounded by a gas of molecular mass $m=6.63\cdot10^{-26}$ kg at temperature $T=300$~K, density $\rho_g=0.011$ kg/m$^3$, and thermal conductivity $\lambda = 18$ mW/(mK). A temperature gradient of $\boldsymbol{\nabla} T=5$~K/mm $\mathbf{e}_z$ is applied such that the heat flux $\mathbf{q}=-\lambda \boldsymbol{\nabla}T$ is directed in negative $z$-direction. Then $\text{Kn} \approx 10$, $u_d\approx 22$~mm/s, $\tau\approx 150$~$\upmu$s, $B_q/(kT) \approx 2.73$ and $B_q^{(\text{eff})}/(kT) \approx 0.45$. For simplicity, the temperature in the vicinity of the translating sphere was assumed not to change appreciably during its motion such that a steady state is attained in the simulation. In figure \ref{fig:Langevin} normalized histograms for the distribution of the azimuthal angle $\theta=\arccos(\mathbf{n}_p\cdot\mathbf{e}_z)$ the particles adopts during the simulations are shown both for a sphere held at a fixed position (left), $\mathbf{u}=0$, and a translating sphere (right), where the particle attains a mean velocity $\langle\mathbf{u}\rangle=\mathbf{u}_d$ according to equation (\ref{eq:uDriftChapmanEnskog}). Note that $\theta=\pi-\tilde{\theta}$. The histograms compare excellently with the angular distributions according to the Boltzmann distribution (\ref{eq:BoltzmannTheta}) with potential (\ref{eq:potential}), both for a stationary sphere (red lines) and for a translating sphere (green dashes).

Note that at a given Kundsen number the necessary temperature gradient for alignment increases rapidly with decreasing particle radius. If for some scale factor $s$ the particle radius $R$ is scaled ${\sim} s$ and the gas density $\rho_g {\sim} s^{-1}$, the Knudsen number remains the same. When simultaneously the temperature gradient (and thus the heat flux) is scaled ${\sim} s^{-3}$, the exponent $B_q/(kT)$ and $B_q^{\text{eff}}/(kT)$ in the Boltzmann distributions of the particle orientation remain the same, while the drift speed $u_d$ scales ${\sim} s^{-2}$ and the timescale $\tau {\sim} s^2$. As illustration, based on the above example and $s=0.1$, an $R=50$~nm Janus particle in a gas with density $\rho_g=0.11$~kg/m$^3$ needs a temperature gradient of $|\boldsymbol{\nabla} T|=5$~K/$\upmu$m to have the same angular distribution as in the example of figure \ref{fig:Langevin} above. 

%%%%%%%%%%%%%%%%%%%%%%%%%%%%%%%%%%%%%%%%%%%%%%%%%%%%%%%%
\section{Free molecular flow between parallel plates: $\ell \gg h$}\label{sec:transferKnInf}

\subsection{Forces and torque on a Janus particle}\label{sec:PhaseSpace_BHS}

When the mean free path $\ell$ of gas molecules is much larger than the separation $h$ of the plates in figure \ref{fig:schematic0}, the Chapman-Enskog expansion cannot be used for the phase space distribution function for the molecules impinging on the particle. Instead, each molecule must be traced back to its position of last diffuse scattering, i.e. the container walls, to infer the corresponding phase space distribution function \cite{Sone_2007}. In particular, we will assume that collisions between molecules can be neglected and only collisions with the walls of the container or the sphere play a role. Without the sphere, the phase space distribution between the parallel diffusely reflecting plates is a superposition of two half-space distributions, 
\begin{equation}\label{eq:binaryHalfspaceDistribution}
\tilde{f}(\mathbf{r}, \tilde{\mathbf{c}}) = f^{(d)}_{l}(\tilde{\mathbf{c}})H(\tilde{\mathbf{c}}\cdot \mathbf{e}_z) + f^{(d)}_{h}( \tilde{\mathbf{c}}) H(-\tilde{\mathbf{c}}\cdot \mathbf{e}_z),
\end{equation}
where $H(x)$ is the Heaviside step function ($H(x)=1$ for $x>0$ and $H(x)=0$ for $x<0$), and 
\begin{equation}
f^{(d)}_{i}(\tilde{\mathbf{c}}) = \nu\frac{2\beta_i^2}{\pi} \exp (-\beta_i \tilde{\mathbf{c}}^2), \qquad \beta_i = \frac{m}{2k T_i} = \frac{1}{\bar{c}_i^2}, \qquad i\in\{l,h\}
\end{equation}
with constant molecule flux density $\nu$ at the walls. Similar to what was done above we have introduced the characteristic molecular velocities $\bar{c}_l=1/\sqrt{\beta_l}$ and $\bar{c}_h=1/\sqrt{\beta_h}$ for gas molecules reflected from the lower and upper wall, respectively. Macroscopically, the gas is at rest and the corresponding molecule density is $n=\int\!\! f(\mathbf{r}, \mathbf{c}) d^3c = \sqrt{\pi}\nu(\sqrt{\beta_l} + \sqrt{\beta_h}) = \sqrt{\pi}\nu(\bar{c}_h^{-1}+\bar{c}_l^{-1})$, while the pressure, temperature and heat flux are \cite{Landau_1983_X} $p=nk\bar{T}=\frac{m}{3}\!\int\! c^2 f(\mathbf{r}, \mathbf{c}) d^3c$, $\bar{T}=\sqrt{T_l T_h}$ and $\mathbf{q}=\frac{m}{2}\! \int\! c^2\,\mathbf{c} f(\mathbf{r}, \mathbf{c}) d^3c=-2\nu k(T_h-T_l) \mathbf{e}_z$, which can be combined to $\mathbf{q}/p=-2\mathbf{e}_z (\bar{c}_h-\bar{c}_l)/\sqrt{\pi}$.

For a stationary, non-rotating sphere and Maxwell-type boundary conditions it can be shown that $\nu(\mathbf{r})=\nu$ is constant on all surfaces \cite{Sone_2007}, and we assume this to hold to a good approximation when the sphere is allowed to move, i.e. we again assume that the phase space distribution of molecules impinging on the sphere is not altered by the presence of the sphere. In the frame of reference of the center of mass of the sphere, moving at velocity $\mathbf{u}$ relative to the gas, the molecular velocities transform as $\mathbf{c}=\tilde{\mathbf{c}}-\mathbf{u}$. As before, the corresponding phase space distribution function in this frame of reference is $f(\mathbf{r}, \mathbf{c}) = \tilde{f}(\mathbf{r}, \tilde{\mathbf{c}}) = \tilde{f}(\mathbf{r}, \mathbf{c}+\mathbf{u})$, which we expand to first order in the particle velocity
\begin{equation}
\begin{split}
f(\mathbf{r}, \mathbf{c}) & \approx  
\nu\frac{2\beta_l^2}{\pi} \exp (-\beta_l \mathbf{c}^2)\left[(1-2\beta_l \mathbf{c}\cdot \mathbf{u})H(\mathbf{c}\cdot \mathbf{e}_z) + (\mathbf{u}\cdot \mathbf{e}_z) \delta (\mathbf{c}\cdot \mathbf{e}_z) \right] \\
& + \nu\frac{2\beta_h^2}{\pi} \exp (-\beta_h \mathbf{c}^2)\left[(1-2\beta_h \mathbf{c}\cdot \mathbf{u})H(-\mathbf{c}\cdot \mathbf{e}_z) - (\mathbf{u}\cdot \mathbf{e}_z) \delta (\mathbf{c}\cdot \mathbf{e}_z) \right],
\end{split}
\end{equation}
where $\delta(x)$ is the Dirac delta function. With this expression and using equation (\ref{eq:partFlux}), the molecule flux density at position $\mathbf{r} = R \mathbf{e}_r$ on the surface of the sphere is 
\begin{align}\label{eq:nu_binaryHalfspace}
\nu(\mathbf{r}) &=\nu\left\{ 1+(\mathbf{u}\cdot\mathbf{n})[\bar{c}_h^{-1}(\pi-\vartheta)+\bar{c}_l^{-1} \vartheta]/\sqrt{\pi} \right\},
\end{align}
where, as shown on the right hand side of figure \ref{fig:schematic0}, the angle $\vartheta$ is the polar angle with $\cos\vartheta=\mathbf{n}\cdot\mathbf{e}_z$ and $0\le\vartheta\le\pi$. Correspondingly, the force density $\mathbf{\Pi}(\mathbf{r})$ on the surface of the sphere is again given by equation (\ref{eq:momFlux}), with
\begin{align}\label{eq:PiIn_binaryHalfspace}
\mathbf{\Pi}_\mathrm{in}(\mathbf{r}) 
&= -\frac{\nu m}{2\sqrt{\pi}}\left\{
\mathbf{n} \left[ \bar{c}_h (\pi-\vartheta) + \bar{c}_l \vartheta \right]
+\mathbf{e}_z \left[ (\bar{c}_h-\bar{c}_l) \sin\vartheta \right]
\right\} 
- \nu m (\mathbf{I}+\mathbf{n} \otimes \mathbf{n})\mathbf{u}\\
\mathbf{\Pi}_\mathrm{out}^\mathrm{(d)}(\mathbf{r}) &= - \nu(\mathbf{r}) m \left[ \frac{\sqrt{\pi} \bar{c}_p}{2} \mathbf{n} + \boldsymbol{\omega}\times\mathbf{r}\right],
\end{align}
where $\bar{c}_p = 1/\sqrt{\beta_p}=\sqrt{2kT_p/m}$ is the characteristic molecular velocity based on the particle temperature. 

The force and torque on the particle are again obtained by integrating $\mathbf{\Pi}(\mathbf{r})$ and $\mathbf{r}\times\mathbf{\Pi}(\mathbf{r})$ over the surface of the sphere. As before, the orientation-dependent integrals over each hemisphere are evaluated in appendix \ref{sec:integrals}. As we have seen in the calculation using the Chapman-Enskog expansion, the terms quadratic in $\mathbf{u}$ and $\boldsymbol{\omega}$ can safely be neglected, as we are interested in particle-velocities far below the typical thermal velocities of gas molecules, and thus we will do so in the following. Under these conditions the force becomes  
\begin{equation}
\begin{split}\label{eq:F_DualHalfSpace}
\mathbf{F} =& -\frac{4}{3} m\nu S_0 \mathbf{u}\left[ 1 + \frac{\pi}{8}\frac{1}{2}(a^+ + a^-) \frac{\bar{c}_p}{2}(\bar{c}_h^{-1}+\bar{c}_l^{-1}) \right] \\
&-\frac{\sqrt{\pi}}{16}m\nu S_0 \left[ 3 (\bar{c}_h - \bar{c}_l)\mathbf{e}_z + (a^+ - a^-) [2\bar{c}_p - (\bar{c}_h + \bar{c}_l)]\mathbf{n}_p \right] \\
&-\frac{1}{4} m\nu S_0 (a^+ - a^-) \bar{c}_p (\bar{c}_h^{-1}-\bar{c}_l^{-1}) \mathbf{u}\cdot\mathbf{N}(\mathbf{n}_p)\\
&-\frac{1}{4} m\nu S_0 (a^+ - a^-) R\boldsymbol{\omega}\times\mathbf{n}_p,
\end{split}
\end{equation}
where the tensor $\mathbf{N}(\mathbf{n}_p)$, linking velocity and particle orientation to the force, can be found in appendix \ref{sec:Integrals_2}. For $\mathbf{u}=\boldsymbol{\omega}=0$ and $a^\pm=1$ the force agrees with the result by Phillips, \cite{Phillips_1972, Gallis_2001}.

Correspondingly, the torque becomes, again neglecting terms quadratic in $\mathbf{u}$ and $\boldsymbol{\omega}$,
\begin{align}\label{eq:M_DualHalfSpace}
\begin{split}
\mathbf{M} =& -\frac{2}{3} R^2 m\nu S_0 \tfrac{1}{2}(a^+ + a^-) \boldsymbol{\omega} \\
&-\frac{1}{4} R m\nu S_0 (a^+ - a^-) \mathbf{n}_p \times \mathbf{u} \\
&-\frac{1}{3\pi\sqrt{\pi}} R m \nu S_0 (\bar{c}_h-\bar{c}_l) (a^+-a^-)  \tau(\theta)\frac{\mathbf{n}_p\times\mathbf{e}_z}{|\mathbf{n}_p\times\mathbf{e}_z|},
\end{split}
\end{align}
with $\theta=\arccos (\mathbf{n}_p\cdot\mathbf{e}_z) \in [0,\pi]$ and (see appendix \ref{sec:Integrals_2})
\begin{align}\label{eq:tau_theta}
\tau(\theta) &= \frac{1}{4} \lvert \cot\theta \rvert \left[(3-\cos(2\theta))E(-\tan^2\theta) - 2K(-\tan^2\theta) \right] \\
&\approx \frac{1}{64} \left(67\sin(\theta)+3\sin(3\theta)\right),\label{eq:tau_theta_approx}
\end{align}
where $K(m)=\int_0^{\pi/2}\left(\sqrt{1-m\sin^2\psi}\right)^{-1}d\psi$ and $E(m)=\int_0^{\pi/2}\left(\sqrt{1-m\sin^2\psi}\right)d\psi$ are the complete elliptic integrals of first and second kind \citep{Abramowitz_1970}. An excellent approximation for $\tau(\theta)$ is given by equation (\ref{eq:tau_theta_approx}), where the coefficients have been selected such that the analytical value $\tau(\pi/2)=1$ is captured, see figure \ref{fig:tau_theta}. We can again define a 'potential' such that $\tau(\theta) = -\partial_\theta {\cal V}_\tau(\theta)$, with ${\cal V}_\tau(\theta) = -\int_{\pi/2}^{\theta}\tau(\theta')d\theta' \approx \frac{1}{64} (67 \cos\theta+\cos (3\theta))$.

We remark that according to eq. (\ref{eq:nu_ChapmanEnskog}) the molecule flux density on a surface at rest in a gas with velocities according to the Chapman-Enskog distribution is $\nu=n/\sqrt{4\pi\beta}=(p/m)\sqrt{\beta/\pi}$. The factors $p\sqrt{\beta/\pi}$ ubiquitous in equations (\ref{eq:F_ChapmanEnskog}) and (\ref{eq:M_ChapmanEnskog}) for the force and torque on the particle in the Chapman-Enskog distribution thus directly correspond to the factors $m\nu$ in equations (\ref{eq:F_DualHalfSpace}) and (\ref{eq:M_DualHalfSpace}), which in particular makes the agreement between the translational and rotational friction factors in the two different formulations transparent. 

Using $\mathbf{q}/p=-2\mathbf{e}_z (\bar{c}_h-\bar{c}_l)/\sqrt{\pi}$, valid for the superposed half-space distributions, the torque can be rewritten as
\begin{align}\label{eq:M_DualHalfSpace_pq}
\mathbf{M} =& -\frac{2}{3} R^2 m\nu S_0 \tfrac{1}{2}(a^+ + a^-) \boldsymbol{\omega}-\frac{1}{4} R m\nu S_0 (a^+ - a^-)\, \mathbf{n}_p \times \left( \mathbf{u} - \frac{10}{3\pi}  \frac{\tau(\theta)}{|\mathbf{n}_p\times\mathbf{e}_z|} \frac{\mathbf{q}}{5p}\right).
\end{align}
Since $10/(3\pi)\approx 1.06$ and, as can be seen in figure \ref{fig:tau_theta}, $\tau(\theta)\approx\sin\theta=|\mathbf{n}_p\times\mathbf{e}_z|$, this agrees well with equation (\ref{eq:M_ChapmanEnskog}) up to first order in $\mathbf{u}$ and $\boldsymbol{\omega}$.

%%%%%%%%%%%%%%%%%%%%%%%%%%%%%%%%%%%%%%%%%%%%%%%%%%%%%%%%
\subsection{Thermophoresis and  orientation of a Janus-sphere}

Just as in the case of a particle in a Chapman-Enskog distribution, its drift velocity is determined by balancing drag and thermophoretic forces. For small temperature differences between the plates, the third term in equation (\ref{eq:F_DualHalfSpace}), linking the drift velocity to the particle orientation, is much smaller than the friction in the first term (note that for each component $|N_{ij}|<1$). The last term in (\ref{eq:F_DualHalfSpace}) will give no net contribution, since again the angular velocity has no preferred direction for a given orientation, as argued for the Chapman-Enskog case. Finally, we assume that the difference in temperature between the two plates is small enough that $\bar{T}=\sqrt{T_lT_h}\approx (T_h+T_l)/2$ and that the temperature of the particle is close to the mean gas temperature, such that the mean force on the particle at steady state is approximately
\begin{equation}
\langle\mathbf{F}\rangle \approx -\frac{4}{3} m\nu S_0 \langle\mathbf{u}\rangle\left[ 1 + \frac{\pi}{8}\frac{1}{2}(a^+ + a^-)) \right]
-\frac{\sqrt{\pi}}{16}m\nu S_0 \left[ 3 (\bar{c}_h - \bar{c}_l)\mathbf{e}_z  \right],
\end{equation}
from which the drift velocity is obtained by solving $\langle\mathbf{F}\rangle=0$ as
\begin{equation}\label{eq:uDriftDualHalfSpace}
\mathbf{u}_d = \frac{45\pi}{128}\frac{\mathbf{q}}{5p} \left[ 1 + \frac{\pi}{16}\left(a^++a^-\right)\right]^{-1},
\end{equation}
where $\mathbf{q}/p=-2\mathbf{e}_z (\bar{c}_h-\bar{c}_l)/\sqrt{\pi}$ was used. Since $45\pi/128\approx 1.10$ this has the same form and agrees well with (\ref{eq:uDriftChapmanEnskog}) obtained using the Chapman-Enskog distribution.

From equation (\ref{eq:M_DualHalfSpace}) or (\ref{eq:M_DualHalfSpace_pq}) the torque on the particle has the form
\begin{equation}
\mathbf{M} = -\hat{\alpha}_\omega \boldsymbol{\omega}
+ \hat{B}_u \mathbf{n}_p\times\mathbf{u}
+ \hat{B}_q\,\tau(\theta)\frac{\mathbf{n}_p\times\mathbf{e}_z}{|\mathbf{n}_p\times\mathbf{e}_z|},
\end{equation}
where the coefficients $\hat{\alpha}_\omega$, $\hat{B}_u$ and $\hat{B}_q$ can directly be read off from equation (\ref{eq:M_DualHalfSpace}).
For a stationary particle, $\mathbf{u}=0$, this again corresponds to a damped dipole in a potential
\begin{equation}\label{eq:pot_binaryHalfSpace}
\hat{V}_q(\theta) = \hat{B}_q {\cal V}_\tau(\theta), \qquad {\cal V}_\tau(\theta) = -\int_{\pi/2}^\theta\tau(\theta')d\theta'\approx \frac{1}{64} (67 \cos\theta+\cos (3\theta)),
\end{equation}
where $\theta$ is the angle between the particle axis, $\mathbf{n}_p$, and the $z$-axis, see figure \ref{fig:schematic0}. Similarly, for a particle moving with the drift velocity, (\ref{eq:uDriftDualHalfSpace}), we can again introduce an effective potential for the particle alignment
\begin{align}
\hat{V}_q^{(\mathrm{eff})}(\theta) &= \hat{B}_q \left({\cal V}_\tau(\theta) - \frac{27\pi^2}{256}\left[ 1 + \frac{\pi}{16}\left(a^++a^-\right)\right]^{-1} \cos\theta \right) \\
&\approx \hat{B}_q {\cal V}_\tau(\theta) \left(1 - \frac{27\pi^2}{256}\left[ 1 + \frac{\pi}{16}\left(a^++a^-\right)\right]^{-1} \right),\label{eq:potEffApprox_binaryHalfSpace}
\end{align}
where in the last line we have used ${\cal V}_\tau(\theta)\approx\cos\theta$. For $a^+=1$ and $a^-=0$, we get $V^{(\mathrm{eff})}(\theta) \approx 0.13 \hat{B}_q {\cal V}_\tau(\theta)$. As in section \ref{sec:thermophoresis_ChapmanEnskog} we can estimate the probability distribution $p(\theta)$ for the particle to have a certain orientation with respect to the plates from a Maxwell-Boltzmann distribution,
\begin{equation}\label{eq:Boltzmann_binaryHalfSpace}
p(\theta) = \hat{{\cal N}} \sin\theta \exp\left(-\hat{V}_q(\theta)/kT\right), \qquad \hat{{\cal N}}^{-1} = \int_0^\pi p(\theta)\,d\theta.
\end{equation}
Here it is not immediately obvious what the correct temperature $T$ in the Maxwell-Boltzmann distribution should be, since the particle is not in thermal equilibrium with any of the walls. However, we will assume that the temperature difference between the walls is small enough that the arithmetic mean temperature $T=(T_l+T_h)/2$ approximately equals the mean temperature $\bar{T}=\sqrt{T_lT_h}$ calculated from the binary half-space distribution between parallel plates, such that the arithmetic mean temperature can be used in (\ref{eq:Boltzmann_binaryHalfSpace}). 

For later reference, we note that using $n = \sqrt{\pi}\nu(\bar{c}_h^{-1}+\bar{c}_l^{-1})$, appropriate for the binary half-space distribution between parallel plates, $\hat{B}_q$ can be written as
\begin{align}\label{eq:dipole_binaryHalfSpace}
\frac{\hat{B}_q}{kT} &= \frac{8nR^3}{3\pi \left(\sqrt{T/T_h}+\sqrt{T/T_l}\right)}\left(\sqrt{\frac{T_h}{T}}-\sqrt{\frac{T_l}{T}}\right) (a^+-a^-) \approx \frac{4}{3\pi}nR^3 \left(\sqrt{\frac{T_h}{T}}-\sqrt{\frac{T_l}{T}}\right) (a^+-a^-).
\end{align}

In the next section we report simulation results for a moving Janus sphere based on the direct simulation Monte Carlo (DSMC) method. This allows us not only to compare with the analytically obtained results of this section, but also to gauge the influence of molecular collisions on the particle translation and alignment at finite Knudsen numbers.

%%%%%%%%%%%%%%%%%%%%%%%%%%%%%%%%%%%%%%%%%%%%%%%%%%%%%%%%
\section{DSMC simulations}\label{sec:DSMC}

We solve the Boltzmann equation by a DSMC method \cite{Bird_1994} in a variant based on the papers \cite{Babovsky_1989, Neunzert_1995}. This is a time splitting method where in a first step one solves the free transport equation (the collisionless Boltzmann equation) for one time step. During the free flow, boundary conditions are taken into account. In the second step (the collision step), the spatially homogenous Boltzmann equation without the transport term is solved. An explicit Euler step is performed. To guarantee positivity of the distribution function during the collision step, a restriction of the time step proportional to the Knudsen number is needed. This means that the method becomes exceedingly expensive for small Knudsen numbers.
  
A computational domain is discretized using a uniform grid size along all axes, resulting in a cubic grid. Since a moving rigid body is immersed in a gas, we divide the computational domain into three sets of grid cells: Gas grid cells completely filled by gas molecules, rigid body grid cells completely covered by the rigid body, and boundary grid cells which are partially filled by gas molecules and partially by the rigid body. We note that due to the motion of the body we have to update the volume of the cells occupied by the gas at every time step for boundary cells. This update can be done efficiently by marking the boundary cells and their neighboring cells near the surface of the rigid body. Only the boundary cells and their neighbors are candidates for boundary cells in the next time step. Some computational effort is necessary to update the volume of boundary cells occupied by the gas. One can take analytical as well as numerical approaches. We refer to \cite{Shrestha_2015} for details on this issue. 
 
We note that we have to apply the reflection boundary condition twice. First it is applied after the free flow of the gas molecules and  second when the rigid body collides with the gas molecules due to its motion. Therefore, the total force exerted by the gas molecules is summed over both steps. Note that this two-step procedure can lead to over-counting of collisions, e.g. when a specific gas molecule reflected off the sphere in the first step collides again in the second step. However, for small Mach number, Ma=$u/c\ll1$, the probability for this to happen is small and using this approximation in \cite{Shrestha_2015} for a sphere undergoing Brownian motion gave exact results for fluctuations in velocity.
 
To determine the motion of the rigid body, the Newton-Euler equations 
\begin{align}
m \frac{d\mathbf{u}}{dt} = \mathbf{F}(t),\qquad I_p\frac{d\boldsymbol{\omega}}{dt} = \mathbf{M}(t),\qquad \frac{d\mathbf{n}_p}{dt} = \boldsymbol{\omega} \times \mathbf{n}_p,
\end{align}
are solved by an explicit Euler scheme. Here the same time step is taken for the Boltzmann and the Newton-Euler equations. The force and the torque on the sphere are computed by summing over individual collisions with the gas molecules. Specifically, the total force and the total torque exerted on the sphere are computed by accumulating the increments of the linear and angular momentum transferred by all the colliding molecules. We again refer to \cite{Shrestha_2015} for details. Note that in order to obtain the correct scale for the fluctuations in the force and torque, the number of simulated molecules must equal the number of gas molecules in the physical situation studied \cite{Hadjiconstantinou_2003}. This can be understood by noting that in a conventional DSMC simulation $s$ independent gas molecules are bundled together in a single DSMC-particle. Thus, during a certain time interval $N_0\pm\sqrt{N_0}$ collisions of gas molecules of mass $m$ with a wall are replaced with $N_0/s\pm\sqrt{N_0/s}$ collisions of DSMC particles of mass $sm$ drawn from the same velocity distribution, resulting in the same mean value for the momentum exchange  $\sim m\bar{c} N_0 = sm\bar{c}\,N_0/s$ with the wall. However, at the same time the fluctuations of the exchanged momentum are increased from $\sim m\bar{c} \sqrt{N_0}$ to $\sim sm\bar{c}\sqrt{N_0/s}=m\bar{c}\sqrt{sN_0}$. Obtaining the correct magnitudes for both mean values and fluctuations of forces in the simulation thus requires $s=1$.

In the following two subsections we consider a freely rotating Janus particle that first has its center of mass held at a fixed position and second is allowed to translate freely. In all cases we consider a monoatomic gas with molecular mass $m = 6.63\cdot10^{-26}$~kg and use a  hard-sphere collision model with diameter $d=3.68\cdot10^{-10}$~m. The initial phase space distribution of the gas is a Maxwellian distribution with an initial temperature, density and mean velocity as parameters. In all cases the initial mean velocity of the gas and the translational and rotational velocity of the rigid body are zero.
 
\subsection{DSMC simulation of rotating Janus sphere}

In the fist case we consider a rotating Janus particle without translation. The computational domain is a cube of side-length 200~nm. A spherical Janus particle of radius 25~nm and density $\rho_p = 1000$ kg/m$^3$ is located at the center of the cube. In order to track the reflection properties on each hemisphere during the motion of the particle, a flag is assigned to boundary particles on the sphere, indicating whether gas molecules are reflected diffusely or specularly from the respective patch of the surface of the Janus particle. The initial temperature of the gas is set to 300~K. The temperatures of the top and bottom wall are equal to 325~K and 275~K, respectively. The temperature on the Janus particle is always 300~K. We apply diffuse reflection boundary conditions at the top and bottom walls of the cube and periodic boundary conditions at the side walls.  

\begin{figure}
\includegraphics[width=6.2cm]{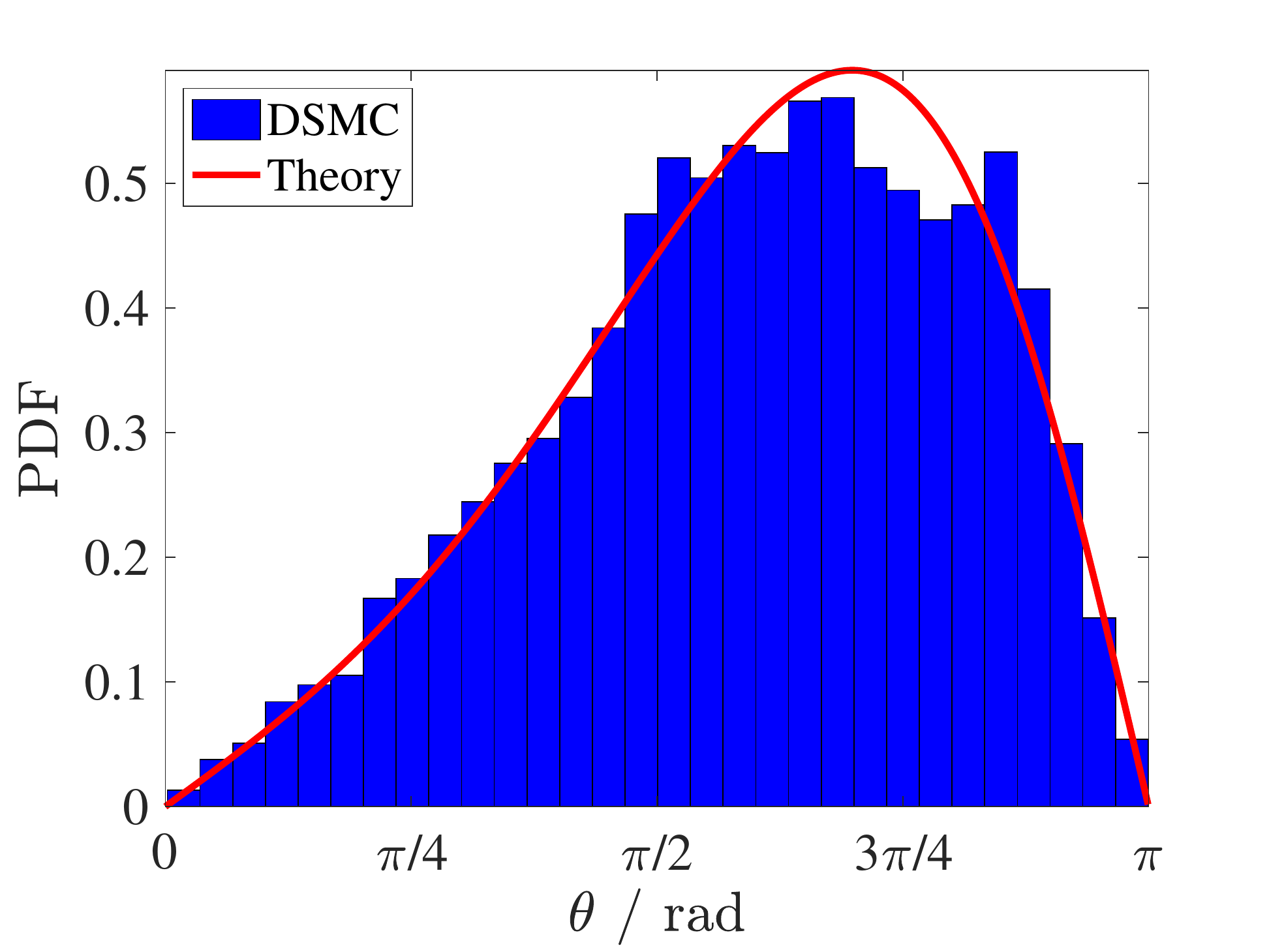} \hspace{0.5cm}
\includegraphics[width=6.2cm]{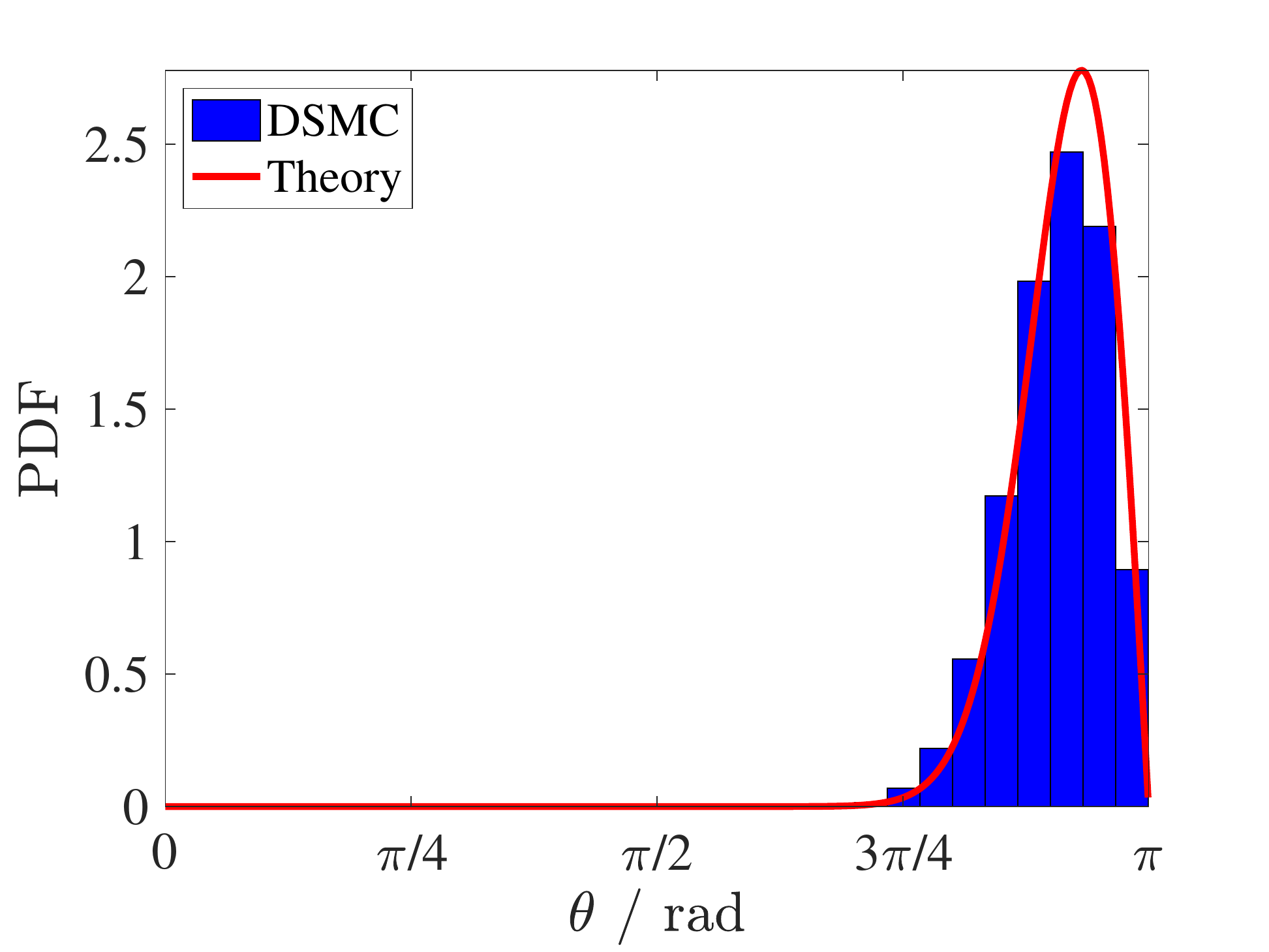}
\caption{\label{fig:angle_rot1} (Color online) Angular distribution for Kn = $\infty$ (collisionless simulation). The left panel corresponds to an initial density $\rho_0$ = 0.1 kg/m$^3$ and the right one to $\rho_0$ = 2.2 kg/m$^3$. 
}
\end{figure}

The domain was discretized by a cubic grid using 10 subdivisions in each direction. We have considered three different values of initial density equal to $\rho_0$ = 0.1~kg/m$^3$, 0.22~kg/m$^3 $ and 2.2~kg/m$^3$, corresponding to Knudsen numbers Kn = 22, 10 and 1, respectively. Since the theoretical results have been derived only for Kn = $\infty$, we have first considered the two cases $\rho_0$ = 0.1~kg/m$^3$ and $\rho_0$ = 2.2~kg/m$^3$ and switched off the intermolecular collisions. For $\rho_0$ = 0.1~kg/m$^3$ the initial number og gas molecules per cell equals 12. Similarly,  $\rho_0$ = 0.22 kg/m$^3$ and 2.2 kg/m$^3$ correspond to 26 and 265 initial molecules per cell, respectively. In figure \ref{fig:angle_rot1} we have plotted the angular distributions according to the analytical results (equations (\ref{eq:Boltzmann_binaryHalfSpace}), (\ref{eq:pot_binaryHalfSpace}) and (\ref{eq:dipole_binaryHalfSpace}) with $T=300$~K) together with the corresponding histograms from the simulations and find both results to be in good agreement.  One can see that for larger density, the Janus particle will be more strongly aligned, with its diffusely reflecting part pointing towards the colder bottom wall. 

\begin{figure}
\includegraphics[width=6.2cm]{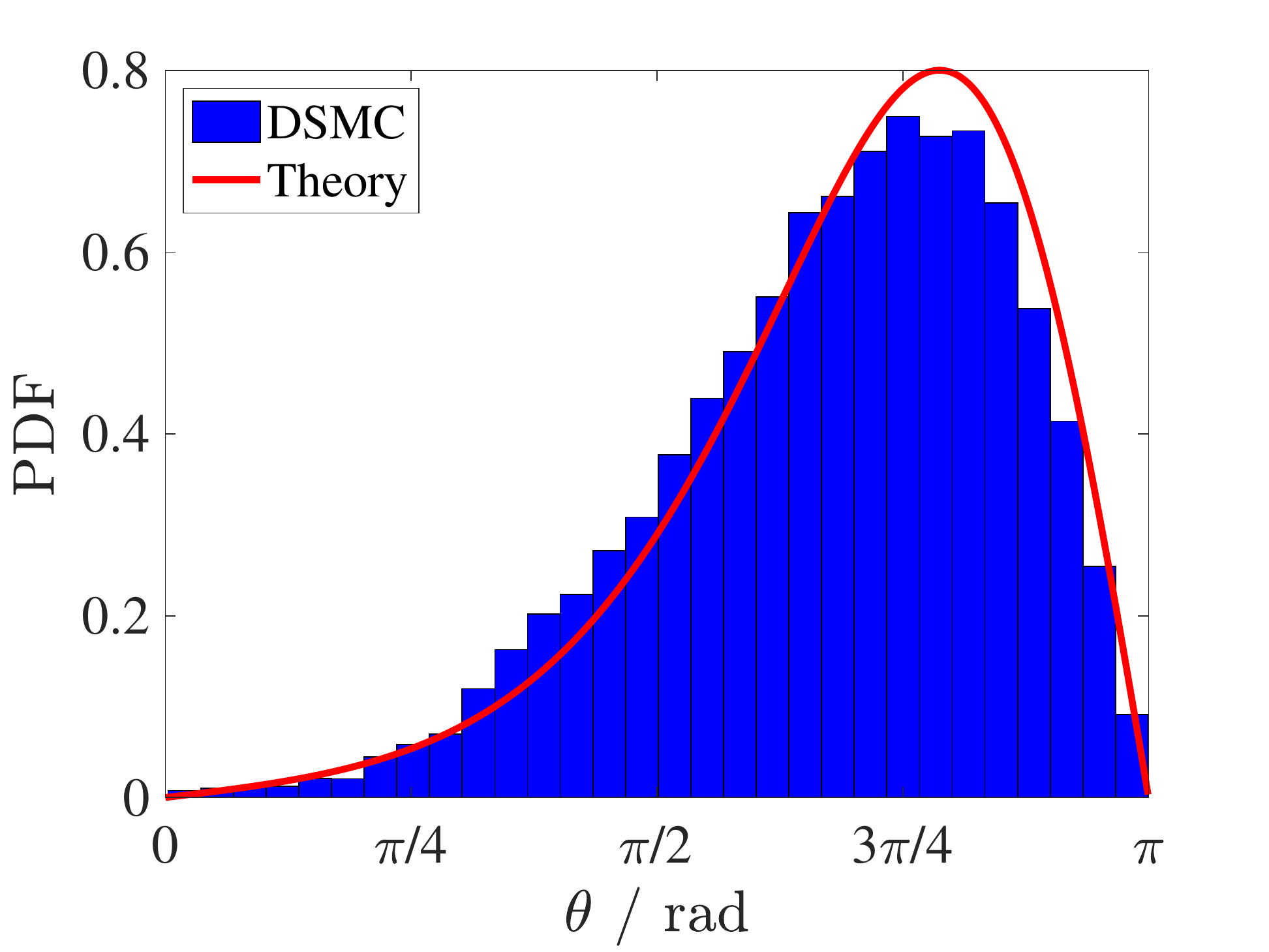} \hspace{0.5cm}
\includegraphics[width=6.2cm]{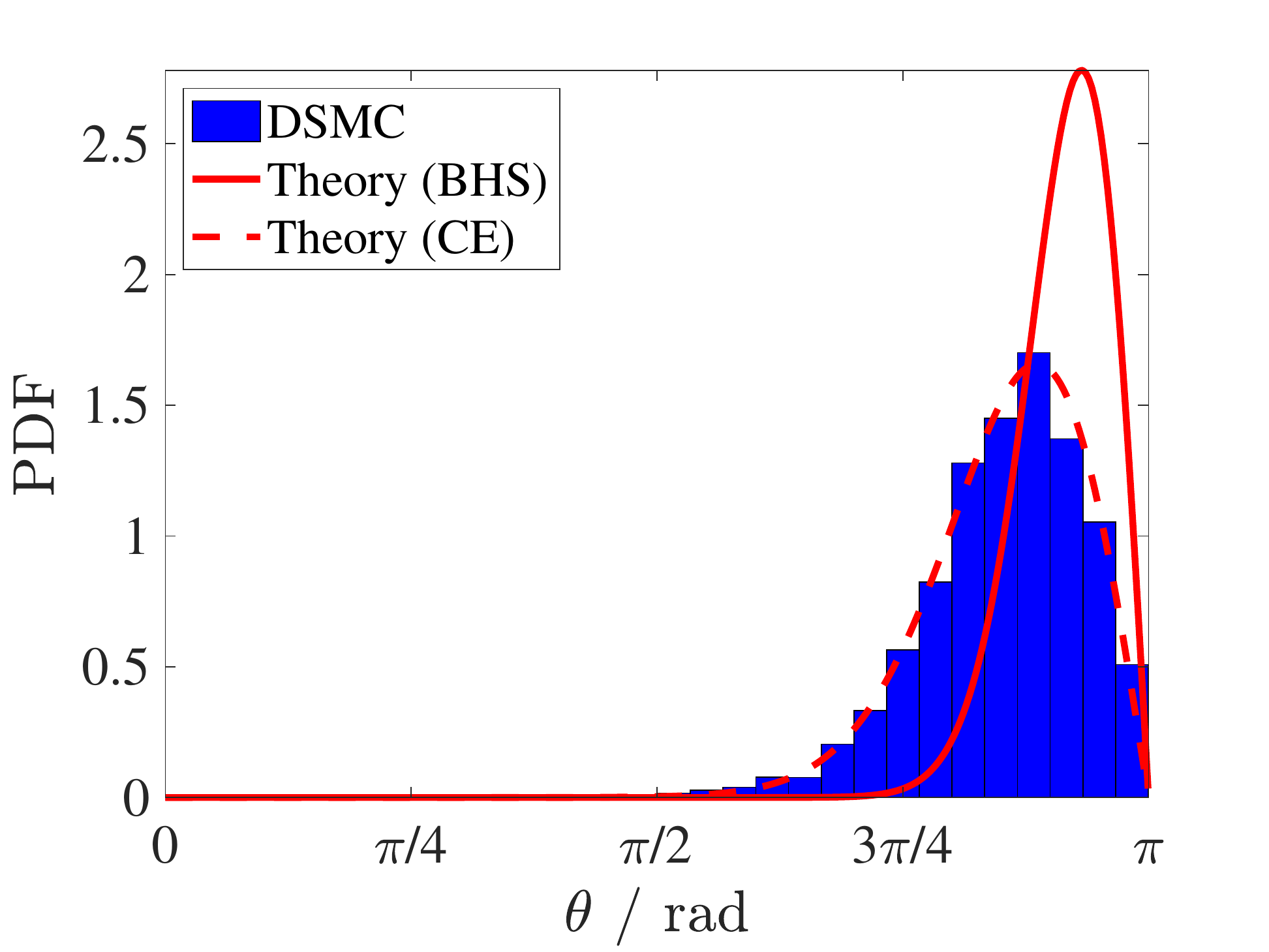}
\caption{\label{fig:angle_rot2} (Color online) Angular distribution for Kn = 10, $\rho_0$ = 0.22 kg/m$^3$ (left) and  Kn = 1, $\rho_0$ = 2.2 kg/m$^3$ (right). 
The solid lines correspond to the theory presented in section \ref{sec:transferKnInf} while the dashed line refers to the results presented in section \ref{sec:ChapmanEnskog}, with a heat flux according to equation (\ref{eq:heatFluxTemperatureJump}).}
\end{figure}

We have further considered the cases with Kn = 10 and 1, i.e. with intermolecular collisions switched on. In figure \ref{fig:angle_rot2} the corresponding angular distributions are plotted together with the analytical results for the collisionless case. One can observe that for Kn = 10 the numerical result is still close to the theoretical values for Kn = $\infty$. However, on the right panel of figure \ref{fig:angle_rot2} one can see that the numerical result of Kn = 1 deviates substantially from the theoretical values for Kn = $\infty$ (labeled BHS). This deviation stems from the fact that in this case the Knudsen number based on the plate distance, $\ell/h$ = 0.25, is relatively small, and it is necessary to take into account the temperature jumps at the upper and lower boundaries of the simulation domain in order to obtain the correct heat flux between the plates. For $\ell/h \lesssim 0.5$ the heat flux between parallel plates at a distance $h$ is approximated well as \cite{Jousten_2008, Sharipov_2007}
\begin{equation}\label{eq:heatFluxTemperatureJump}
q = q_0 \left(1+\frac{4\zeta_T}{\sqrt{\pi}}\frac{\ell}{h}\right)^{-1},
\end{equation}
where $\zeta_T = 1.954$ is the rarefaction parameter, and $q_0=-\lambda (T_h-T_l) / h$ is the continuum heat flux with thermal conductivity $\lambda \approx \frac{15}{4\sqrt{\pi}} k n \bar{c} \ell$, appropriate for a hard sphere gas with $\ell^{-1}=\sqrt{2}\pi d^2 n$ and $\bar{c}=\sqrt{2kT/m}$. Corresondingly, the prediction based on the Chapman-Enskog distribution, (\ref{eq:BoltzmannTheta}), (\ref{eq:potential}) with $T$ = 300 K and using (\ref{eq:heatFluxTemperatureJump}) for the heat flux, leads to excellent agreement between the theory and the simulations, as indicated by the dashed line labeled CE on the right panel of figure \ref{fig:angle_rot2}.
 
%%%%%%%%%%%%%%%%%%%%%%%%%

%%%%%%%%%%%%%%%%%%%%%%%%%%%%%%%
\subsection{DSMC simulation of a rotating and translating Janus sphere}
In this subsection we considered a freely rotating and translating Janus particle with a radius of $25$~nm. All problem parameters are unchanged with respect to the previous case, apart from the temperatures of the top and bottom walls which are now 337.5~K and 262.5~K, respectively, and the size of the simulation box. In addition to rotation we have translational motion in this case, since due to the thermophoretic force the particle moves from the top to the bottom.  It is not straightfoward to simulate a long period of time, since the moving sphere crosses the boundaries. Therefore, we have considered a geometry of size $ \Delta x =$ 200~nm, $\Delta y =$ 200~nm, $\Delta z =$ 300~nm, again discretized with a cubic grid with cells of side lengths of 20~nm. The initial center of the sphere is at $z$ = 200~nm and centered in the $x-y$-slice cutting through the box, where the origin of the coordinate system is located at the bottom wall. When the center of the sphere crosses the plane $z_c$ = 100~nm, we translate it to $z$ = 200~nm at fixed $x$- and $y$-coordinates. Simultaneously, gas molecules within the updated volume of the sphere are translated to the sphere's original position. When the sphere crosses the side walls, we apply periodic boundary conditions. 
 
\begin{figure}
	\includegraphics[width=6.2cm]{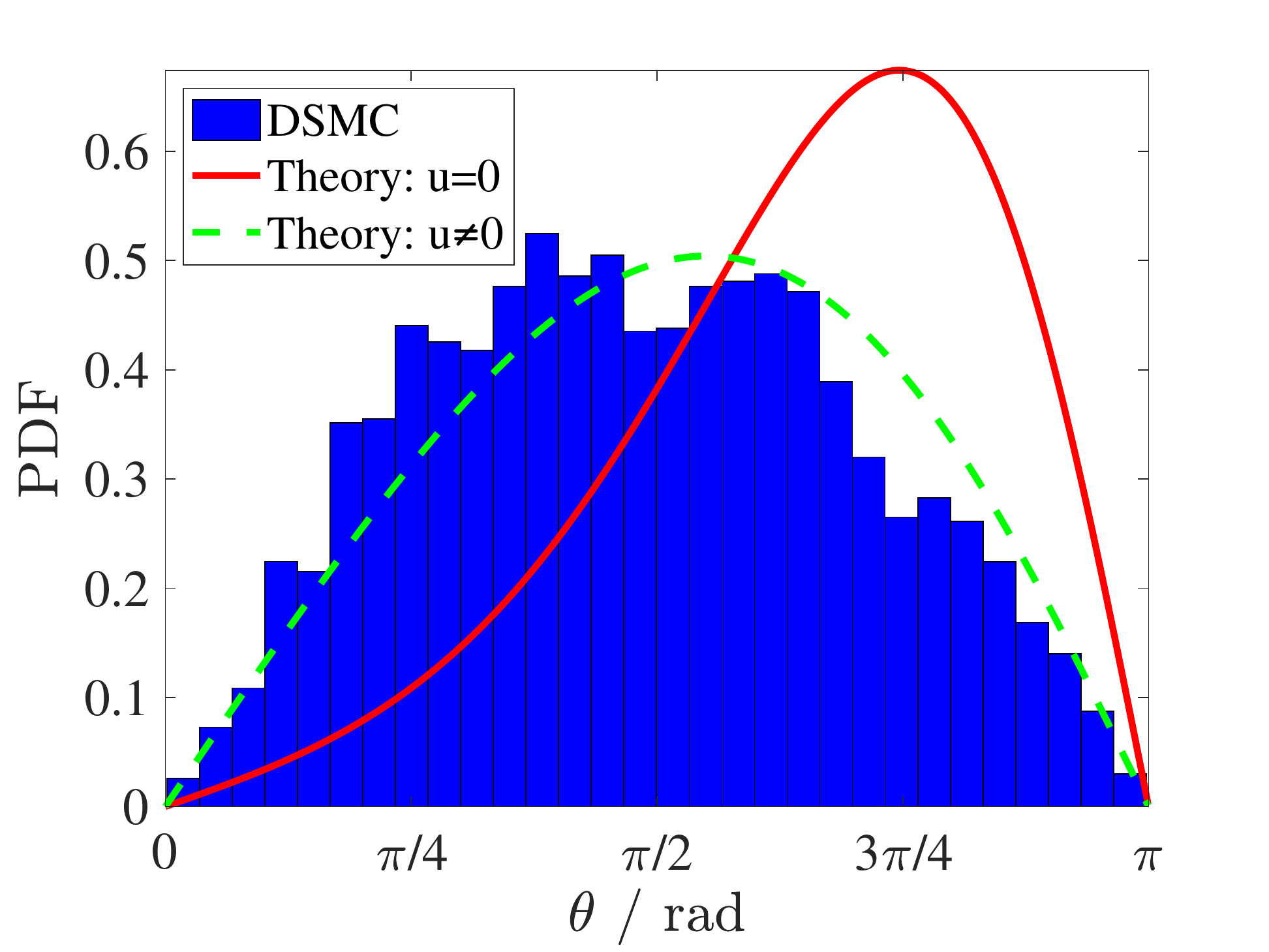} \hspace{-0.6cm}
	\includegraphics[width=6.2cm]{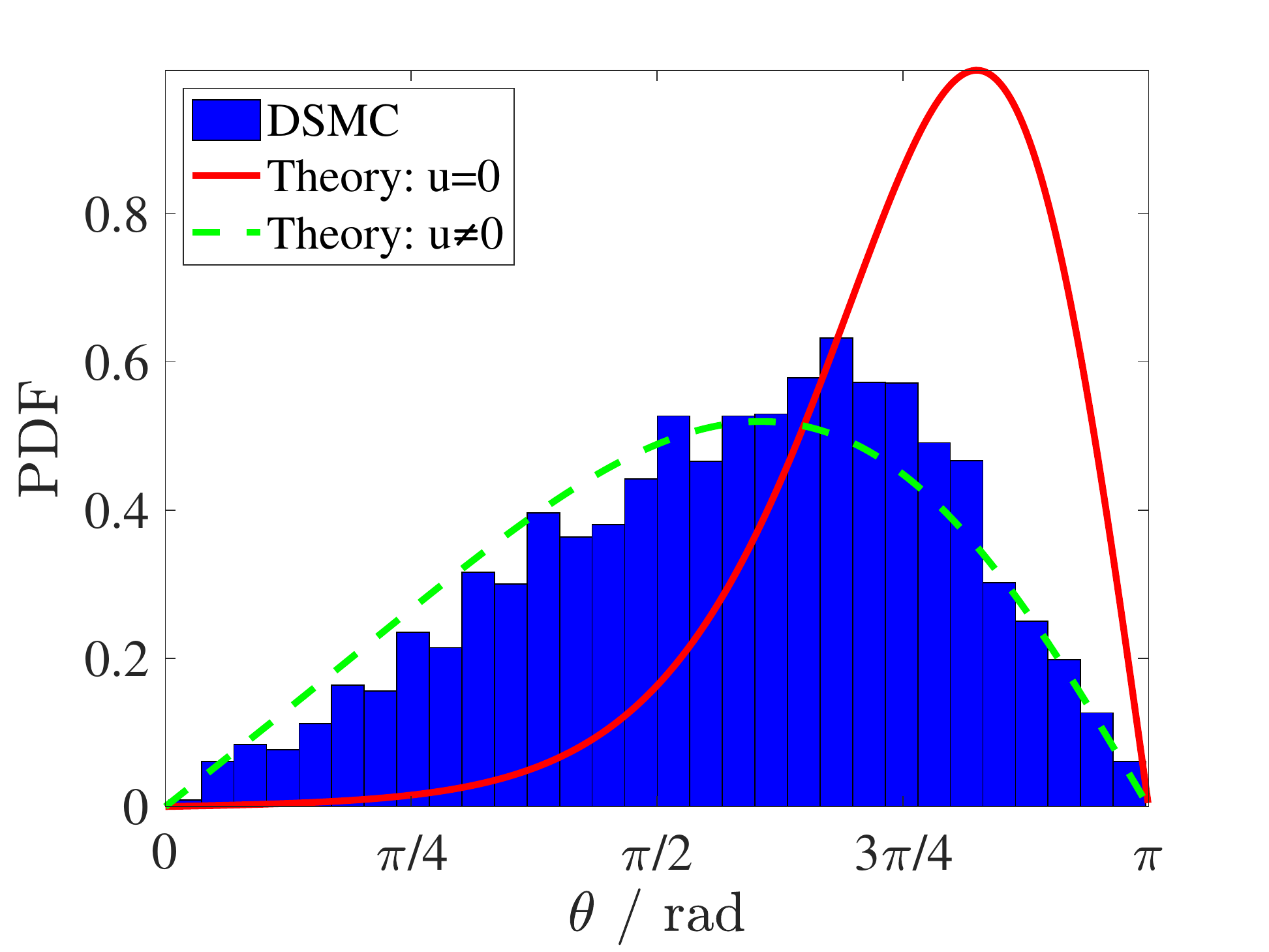} \hspace{-0.6cm}
	\includegraphics[width=6.2cm]{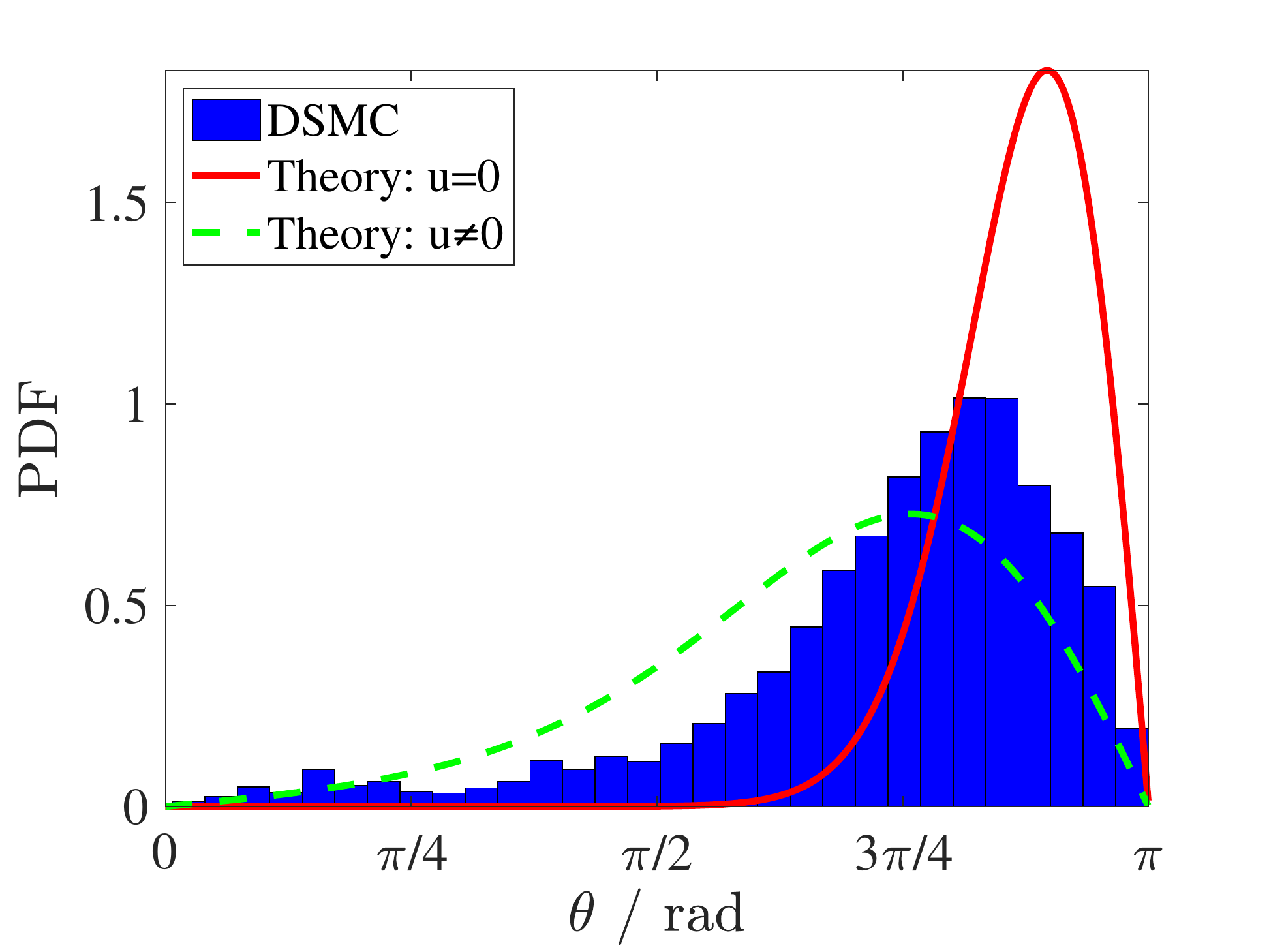}
	\caption{\label{fig:angle_trans} (Color online) The angle distribution of a rotating and translating Janus sphere for Kn = $\infty$, $\rho_0$ = 0.1 kg/m$^3$ (left), Kn = 10, $\rho_0$ = 0.22 kg/m$^3$ (center) and Kn = 1, $\rho_0$ = 2.2 kg/m$^3$ (right). 
		The theoretical predictions are according to section \ref{sec:transferKnInf} for Kn = $\infty$ and 10, while section \ref{sec:ChapmanEnskog} with a heat flux according to equation (\ref{eq:heatFluxTemperatureJump}) was used for Kn = 1. 
	}
\end{figure}

In figure \ref{fig:angle_trans} we have plotted the angular distributions according to the theoretical prediction against the simulated ones for Kn = $\infty$, 10 and 1.  For Kn = $\infty$ and 10 the theoretical prediction is based on the binary half space distribution, equations (\ref{eq:Boltzmann_binaryHalfSpace}), (\ref{eq:potEffApprox_binaryHalfSpace}) and (\ref{eq:dipole_binaryHalfSpace}), while for Kn = 1 it is based on Chapman-Enskog distribution using equations (\ref{eq:BoltzmannTheta}), (\ref{eq:potential}), (\ref{eq:potEffCE}) with T = 300 K, and a heat flux according to (\ref{eq:heatFluxTemperatureJump}). We again find reasonable agreement between the theoretical prediction and the simulation, even for Kn = 1, where deviations between the theoretical prediction, valid for Kn = $\infty$, and simulations are expected.

Finally, we have extracted the Janus particle's mean velocity in $z$-direction. For Kn = $\infty$ the average numerical value for the drift velocity of 9.5~m/s is very close to theoretical one, 9.2~m/s from eq. (\ref{eq:uDriftDualHalfSpace}). However, for Kn = 10 the mean particle drift velocity of 7.1~m/s from the simulation already deviates from the theoretical value at Kn = $\infty$. For Kn = 1 the simulated drift velocity is 2.4 m/s, while eq. (\ref{eq:uDriftChapmanEnskog}), with a heat flux according to eq. (\ref{eq:heatFluxTemperatureJump}), predicts a drift velocity of 3.0 m/s under these conditions. Again, it needs to be kept in mind that the prediction is strictly valid only in the limit Kn $\gg$ 1, so a deviation between the theoretical prediction and the simulation is expected.

%%%%%%%%%%%%%%%%%%%%%%%%%%%%%%%%%%%%%%%%%%%%%%%%%%%%%%%%
\section{Conclusion and Outlook}\label{sec:ConclusionOutlook}

In this paper we have investigated the translation and rotation of a Janus particle in a gas at rarefied conditions under the influence of a temperature gradient. Janus particles with different reflection properties on their respective hemispheres were considered. The main results obtained are expressions for the probability density of the angle between the particle's symmetry axis and the heat flux, both for a stationary particle and a translating one. It is found that the particle's alignment is severely impacted by the thermophoretic translation of the particle. This is due to the fact that thermophoretic forces align the particle with its diffuse side to the colder region, while simultaneously causing the translation of the particle in this direction. During the resulting thermophoretic motion of the aligned particle, the increased number of gas molecules impinging on the windward diffuse side results in an adverse torque, reducing the alignment. The Brownian motion of the particles' orientation can be considered as a random walk in a potential with strength proportional to the heat flux in the gas. For a freely translating sphere this potential is reduced to less than 1/5 of the potential a non-translating sphere experiences under the same conditions. 

The scenario considered may be viewed as characteristic for the deposition of small particles with a dipolar structure on a cold surface. A large temperature gradient can help depositing the particles in an oriented manner. With that goal in mind it will be advantageous to reduce the drift velocity of the particles to be deposited. One possibility to achieve this is to deposit charged particles in an additionally applied electric field. For microparticles gravity can partially compensate the thermophoretic force and reduce the drift velocity or even allow particle deposition against the thermophoretic force. However, in this case large density inhomogeneities within the Janus-particle, for example originating from a metallic coating of a polymer particle, may influence the alignment, as gravity may exert an additional torque on the particle. To illustrate, the characteristic gravitational energy scale of orientation for a spherical Janus particle with a density mismatch $\Delta\rho$ = 5000~kg/m$^3$ between its hemispheres is $E_g = \pi R^4 g \Delta\rho/4 \sim 10^{-3} kT$ for $R=100$~nm, $g = 9.8$~m/s$^2$ and $T=300$~K, but grows with the forth power of the particle radius.

We would like to mention that the results were so far obtained without taking the thermal inertia of the particle into account. Since the thermoporetic particle motion is towards colder regions, its surface will be warmer than the surrounding gas phase (the timescales for thermal equilibration and alignment are estimated in section \ref{sec:thermalTimescales} of the appendix). As we noted at the end of section \ref{sec:PhaseSpace_ChapmanEnskog}, the temperature profile on the particle surface does not result in an altered torque on the particle, however, it contributes to a force slowing the particle down (2nd line in eq. (\ref{eq:F_ChapmanEnskog})). As this slower motion results in a reduced torque on the diffuse surface of the sphere, this effect results in better alignment than predicted by the present estimate.

Finally, we mention that particles of complex shape may also align with the temperature gradient, as considered by \citet{Shrestha_2015}, where a chiral particle with a diffuse-reflection boundary condition at its surface in a thermal gradient was considered. This indicates that the results of this paper are of relevance for a broader class of problems. How the alignment of a particle in a temperature gradient depends on its shape, i.e. which particle shapes result in alingment, is an open problem that could define a quite extensive arena for future studies.

%%%%%%%%%%%%%%%%%%%%%%%%%%%%%%%%%%%%%%%%%%%%%%%%%%%%%%%%%%%%%%%%%%%
\begin{acknowledgments}
Financial support by the DFG (Deutsche Forschungsgemeinschaft) under grant numbers HA 2696/41-1 and KL 1105/27-1 is gratefully acknowledged.
\end{acknowledgments}

%%%%%%%%%%%%%%%%%%%%%%% Bibliography %%%%%%%%%%%%%%%%%%%%%%%
% Create the reference section using BibTeX:
%\bibliographystyle{abbrvnat} %Author's initials: https://tex.stackexchange.com/questions/36660/only-authors-initials-in-bibtex-natbib-using-named-style
%\bibliographystyle{abbrvunsrtnat} %https://tex.stackexchange.com/questions/106152/how-can-i-modify-abbrvnat-bst-to-have-unsorted-reference-list-using-natbib
%\bibliography{litRarefiedPhoresis}

%%%%%%%%%%%%%%%%%%%%%%%% Appendix %%%%%%%%%%%%%%%%%%%%%%%%

%\balancecolsandclearpage
%\clearpage

\appendix
\section*{Appendix}
%%%%%%%%%%%%%%%%%%%%%%%%%%%%%%%%%%%%%%%%%%%%%%%%%%%%%%%%
\section{Surface integrals\label{sec:integrals}}

In a Cartesian coordinate system with origin in the centre of the sphere and the orthogonal unit vectors $\mathbf{e}_x$, $\mathbf{e}_y$ and $\mathbf{e}_z$, a parametrisation of the unit normal vectors on the surface of the sphere reads, c.f. figure \ref{fig:schematic0}.
\begin{align}
\mathbf{n} \equiv \mathbf{n}(\vartheta,\varphi) = \sin\vartheta \cos\varphi \, \mathbf{e}_x + \sin\vartheta \sin\varphi\,\mathbf{e}_y +  \cos\vartheta \, \mathbf{e}_z, \qquad 0\le\vartheta\le\pi,\quad 0\le\varphi<2\pi.
\end{align}
The orientation of the Janus sphere is characterised by the orientation vector, $\mathbf{n}_p$, normal to the equatorial plane of the sphere. This separates the points $\mathbf{r}=\mathbf{r}(\vartheta,\varphi)=R\,\mathbf{n}(\vartheta,\varphi)$ on the surface of the sphere into the sets $S^{+} = \{\mathbf{r}=R\mathbf{n}\,|\,\mathbf{r}\cdot\mathbf{n}_p > 0\}$ and $S^{-} = \{\mathbf{r}=R\mathbf{n}\,|\,\mathbf{r}\cdot\mathbf{n}_p < 0\}$ on the upper and lower hemisphere, respectively. For $\mathbf{n}_p$ we use the representation
\begin{align}
\mathbf{n}_p = \sin\theta \cos\phi \, \mathbf{e}_x + \sin\theta \sin\phi\,\mathbf{e}_y +  \cos\theta \, \mathbf{e}_z, \qquad 0\le\theta\le\pi,\quad 0\le\phi<2\pi.
\end{align}

Below we will use the abbreviations $dS = R^2\sin\vartheta \, d\vartheta \, d\varphi$ for the infinitesimal surface element on the sphere and $S_0 = 4\pi R^2$ for its surface area.

\subsection{Integrals: Janus sphere in a Chapman-Enskog distribution\label{sec:Integrals_1}}

For evaluating the integrals of $\Pi(\mathbf{r})$ and $\mathbf{r}\times\Pi(\mathbf{r})$ over the surface, integrals of tensor products of the normal vector over a hemisphere are needed. We will here use a coordinate representation such that $n_i=(\mathbf{n}\cdot\mathbf{e}_i)$ is the $i$th component of the normal vector $\mathbf{n}$, the tensor $\mathbf{n}\otimes\mathbf{n}$ has the representation $(\mathbf{n}\otimes\mathbf{n})_{ij}=n_i n_j$ and so forth. 

Since in this case the only distinguished vector is the orientation vector $\mathbf{n}_p$, normal to the equatorial plane of the Janus particle, we can without loss of generality choose to perform these integrals in a Cartesian coordinate system where the particle's axis of symmetry coincides with the $z$-axis. Thus, in this case the integrals over the upper and lower hemispheres, $S^{+}$ and $S^{-}$, are restricted to $0\le\vartheta\le\pi/2$ and $\pi/2\le\vartheta\le\pi$, respectively. With these preliminaries the integrals can be directly performed, and we obtain for the required integrals over the upper hemisphere, $S^{+}$, 
\begin{align}
{\cal S}^{sph,+} &= \int_{S^{+}} dS = \tfrac{1}{2} S_0 \\
{\cal S}^{sph,+}_i &= \int_{S^{+}} n_i dS = \tfrac{1}{4} S_0 n_{p,i} \\
{\cal S}^{sph,+}_{ij} &= \int_{S^{+}} n_i n_j dS = \tfrac{1}{6} S_0 \delta_{ij}\\
{\cal S}^{sph,+}_{ijk} &= \int_{S^{+}} n_i n_j n_k dS = \tfrac{1}{16} S_0 [\delta_{ij} n_{p,k} + \delta_{jk} n_{p,i} + \delta_{ki} n_{p,j} - n_{p,i} n_{p,j} n_{p,k}]
\end{align}
where $\delta_{ij}=\mathbf{e}_i\cdot\mathbf{e}_j$ is the Kronecker delta and $n_{p,i}=\mathbf{n}_p\cdot\mathbf{e}_i$. With this notation, these expressions are valid for any orientation.

The corresponding integrals over the lower hemisphere are inferred from those of the upper hemisphere by symmetry,
\begin{align}
&{\cal S}^{sph,-} = {\cal S}^{sph,+}, \quad {\cal S}^{sph,-}_{\mu\nu} = {\cal S}^{sph,+}_{\mu\nu}\\
&{\cal S}^{sph,-}_\mu = -{\cal S}^{sph,+}_\mu, \quad {\cal S}^{sph,-}_{\mu\nu\sigma} = -{\cal S}^{sph,+}_{\mu\nu\sigma}.
\end{align}

\subsection{Integrals: Janus sphere in a binary half-space distribution \label{sec:Integrals_2}}
In the case of a Janus sphere in a binary half-space distribution, (\ref{eq:binaryHalfspaceDistribution}), both the orientation of the particle, $\mathbf{n}_p$, as well as the polar angle $\vartheta=\arccos(\mathbf{n}_p\cdot\mathbf{e}_z)$ on the surface of the sphere enter the integrals over the force and torque densities, c.f. equations (\ref{eq:nu_binaryHalfspace}) and (\ref{eq:PiIn_binaryHalfspace}). Without loss of generality, the integrals can be evaluated for $\phi=0$ and transformed after evaluation, using $\mathbf{R}_z(\phi)$, the matrix representing a rotation around the $z$-axis by an angle $\phi$ with Cartesian representation
\begin{align}
\mathbf{R}_z(\phi) = \left(\begin{array}{ccc}
\cos\phi & -\sin\phi & 0 \\ 
\sin\phi & \cos\phi & 0 \\ 
0 & 0 & 1
\end{array}\right).
\end{align}

The required integrals over the upper and lower hemispheres, $S^{+}$ and $S^{-}$ are 
\begin{align}
\int_{S^\pm}[a(\pi-\vartheta)+b\vartheta] \mathbf{n} \, dS  &=\int_{S^\pm}[\tfrac{1}{2}(a+b)\pi + \tfrac{1}{2}(a-b)(\pi-2\vartheta)] \mathbf{n} \, dS = 4\pi R^2 \frac{\pi}{8} \left[ \pm (a+b)\mathbf{n}_p +\frac{1}{2}(a-b)\mathbf{e}_z  \right] \\
\int_{S^\pm}[(\pi-2\vartheta) \mathbf{n}\, dS  &= 4\pi R^2 \frac{\pi}{8} \mathbf{e}_z,
	\qquad \int_{S^\pm}\mathbf{n}\, dS  = \pm 4\pi R^2 \frac{1}{4} \mathbf{n}_p 
	\qquad \int_{S^\pm}\mathbf{n}\otimes\mathbf{n}\, dS  = 4\pi R^2 \frac{1}{6} \mathbf{I}\\ %\mathds{1} \\
\int_{S^\pm}[(\pi-2\vartheta) \mathbf{n}\otimes\mathbf{n} \, dS  &= \pm 4\pi R^2 \mathbf{R}_z(\phi) \mathbf{N}(\theta,0) \mathbf{R}_z(-\phi),
	\qquad \mathbf{N}(\mathbf{n}_P)\equiv\mathbf{N}(\theta,\phi)=\mathbf{R}_z(\phi) \mathbf{N}(\theta,0) \mathbf{R}_z(-\phi)\\
N_{11}(\theta,0) &\approx \frac{1}{4\pi} \frac{2}{9} \left[(3\pi-7)\pi\cos\theta - 7(1-\cos^2\theta)\cos\theta \right]\\
N_{22}(\theta,0) &\approx \frac{1}{4\pi} \frac{2}{9} \left[(3\pi-7)\pi\cos\theta \right]\\
N_{33}(\theta,0) &\approx \frac{1}{4\pi} \frac{2}{9} \left[(3\pi-4)\pi\cos\theta - \frac{19}{2}(1-\cos^2\theta)\cos\theta \right]\\
N_{13}(\theta,0) &= \frac{1}{4\pi} \frac{4}{9} \cot\theta\left[ -4\cos(2\theta)E(-\tan^2\theta)+(1+3\cos(2\theta))K(-\tan^2\theta)\right] \approx \frac{1}{4\pi} \frac{16}{9} \sin^3\theta\\
N_{12}(\theta,0) &= N_{21}(\theta,0) = N_{23}(\theta,0) = N_{32}(\theta,0) = 0\\
\int_{S^\pm} \sin\vartheta \, dS & = 4\pi R^2 \frac{\pi}{8}, \qquad
	\int_{S^\pm} \sin\vartheta \cos\vartheta \, \mathbf{n} \, dS = 4\pi R^2 \frac{\pi}{32} \mathbf{e}_z
\end{align}
 By symmetry, the last two integrals must be independent of $\theta$, since under reflection at the center of the sphere, $(\vartheta,\varphi)\to (\pi-\vartheta,\pi+\varphi)$, both $\sin\vartheta$ and $\cos\vartheta\,\mathbf{n}$ are symmetric, and it suffices to evaluate them for $\theta=0$. Similar arguments can be used for the first integral; in particular for $a=b$ the only distinguished vector is $\mathbf{n}_p$, and without loss of generality $\theta=0$ can be considered. We were not able to obtain analytical expressions for the diagonal components of $\mathbf{N}(\theta,0)$ and have instead reported expressions that fit well to numerical evaluations of the corresponding integrals.

Obtaining the net torque on the particle requires evaluating the integral $\int_{S^+} \sin\theta \; (\mathbf{e}_z \times \mathbf{e}_r)\cdot\mathbf{e}_y\, dS$, which can be done as follows for $0\le \theta \le \pi/2$ (here $H(x)$ is the Heaviside step function introduced in equation (\ref{eq:binaryHalfspaceDistribution}), and the normalization is chosen such that $\tau(\pi/2)=1$):
\begin{align}
\frac{8}{3} \tau(\theta) &= \frac{1}{R^2}\int_{S^+} \sin\theta \; (\mathbf{e}_z \times \mathbf{e}_r)\cdot\mathbf{e}_y dS
= \frac{1}{R^2}\int_{S} H(\mathbf{r}\cdot\mathbf{n}_p) \sin\theta \; (\mathbf{e}_z \times \mathbf{e}_r)\cdot\mathbf{e}_y dS\\
&= \int_0^\pi \int_0^{2\pi} H(\sin\theta \cos\varphi \sin\vartheta +\cos\theta \cos\vartheta) \sin\vartheta (-\sin\vartheta \cos\varphi) \sin\vartheta d\varphi d\vartheta\\
&= -4\int_{\pi/2}^{\pi/2+\theta} \int_0^{\pi/2} H(\sin\theta \cos\varphi \sin\vartheta +\cos\theta \cos\vartheta) \sin^3\vartheta \cos\varphi d\varphi d\vartheta\\
&= -4\int_0^\theta \int_0^{\pi/2} H(-\cos\theta \sin\vartheta + \sin\theta \cos\vartheta \cos\varphi) \cos^3\vartheta \cos\varphi d\varphi d\vartheta\\
&= -4 \int_0^{\pi/2} \int_0^{\arctan(\tan\theta \cos\varphi)}  \cos^3\vartheta \cos\varphi d\vartheta d\varphi\\
&=-\frac{2}{3} \cot \theta \left[ 2 K\left(-\tan^2 \theta \right) - [3 - \cos (2 \theta )] E\left(-\tan^2 \theta \right)\right]
\end{align}
where $K(m)=\int_0^{\pi/2}\left(\sqrt{1-m\sin^2\psi}\right)^{-1}d\psi$ and $E(m)=\int_0^{\pi/2}\left(\sqrt{1-m\sin^2\psi}\right)d\psi$ are the complete elliptic integrals of the first and second kind \citep{Abramowitz_1970}. In the second step we used the fact that the integral vanishes for $\theta=0$, such that for $0\le \theta \le \pi/2$ the total integral can be found by adding the integral over the region $\pi/2 \le \vartheta \le \pi/2+\theta$ and subtracting the corresponding region on the other side of the sphere. Using the symmetry of the integrand and restricting the integral over $\varphi$ to only $0\le\varphi\le\pi/2$ finally yields the factor 4 in front of the integral over a reduced region. For $\pi/2 \le \theta \le \pi$, the relation $\tau(\theta)=\tau(\pi-\theta)$ has to be used; the expression for $\tau(\theta)$ is preserved under this symmetry when replacing $\cot\theta$ by its absolute value. 

\begin{figure}
	\includegraphics[width=8cm]{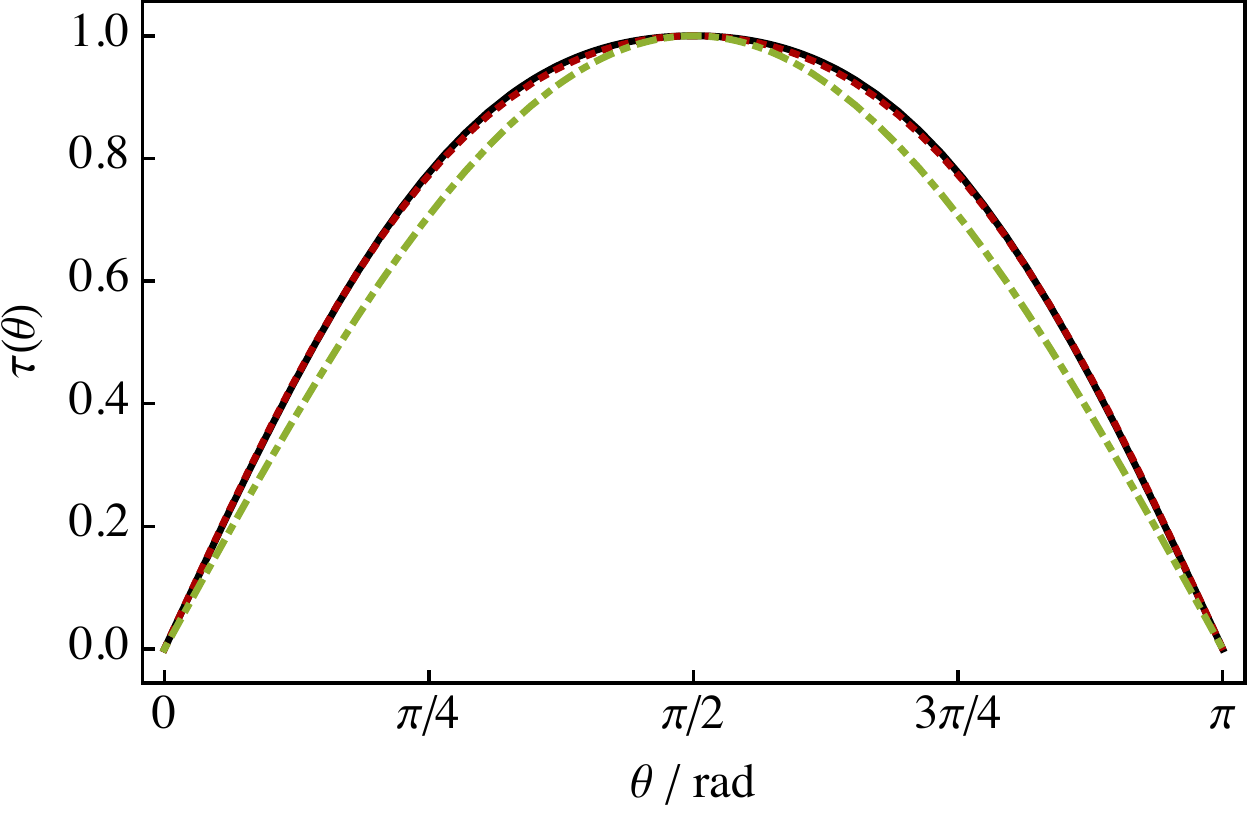} \hspace{0.5cm}
	\includegraphics[width=8cm]{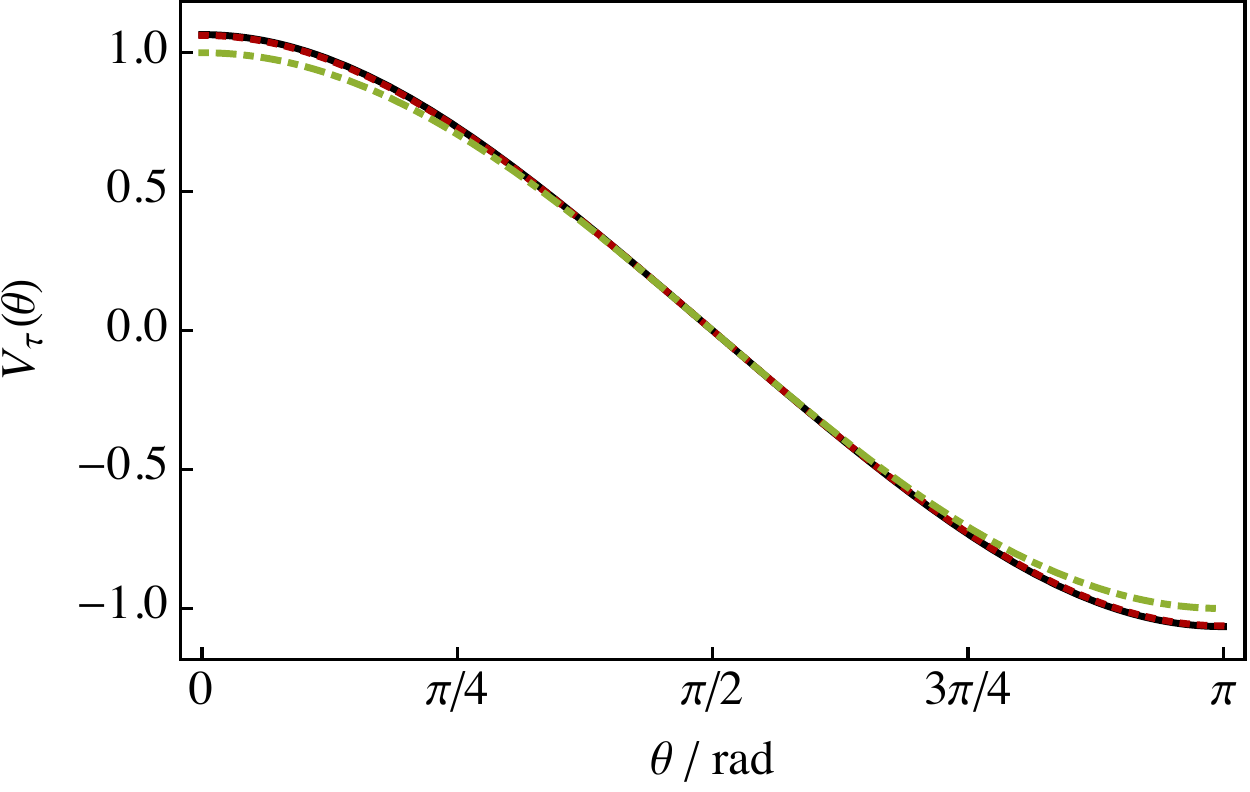}
	\caption{\label{fig:tau_theta} (Color online) Left: Function $\tau(\theta)$ from equation (\ref{eq:tau_theta}) (solid black line) and its approximation by equation (\ref{eq:tau_theta_approx}) (red dotted line). For comparison we have also plotted $\sin\theta$, the functional dependence of the torque on the orientation found in the Chapman-Enskog case (green dash-dotted line). Right: The same for ${\cal V}_\tau(\theta) = -\int_{\pi/2}^{\theta}\tau(\theta')d\theta'$.
	}
\end{figure}

An excellent approximation to $\tau(\theta)$ in terms of trigonometric functions is given by equation (\ref{eq:tau_theta_approx}), as shown in figure \ref{fig:tau_theta}. The agreement between the approximate expressions for the diagonal parts of the tensor $N_{ij}$ introduced above and their numerically obtained values is of equal quality.

%%%%%%%%%%%%%%%%%%%%%%%%%%%%%%%%%%%%%%%%%%%%%%%%%%%%%%%%
\section{Numerical implementation of the Newton-Euler-Langevin equations by Euler discretisation\label{sec:nondimEuler}}

The Newton-Euler-Langevin equations (\ref{eq:LangevinU}-\ref{eq:LangevinN}) are non-dimensionalised using a timescale $\tau=m_p/(S_0p\sqrt{\beta/\pi})$, a velocity scale $U_0=R/\tau$ and a frequency scale $\Omega_0=1/\tau$ such that $\underaccent{\tilde}{t}=t/\tau$, $\underaccent{\tilde}{\mathbf{u}}=\mathbf{u}/U_0$ and $\underaccent{\tilde}{\boldsymbol{\omega}}=\boldsymbol{\omega}/\Omega_0$ become the dimensionless time, velocity and angular frequency.
Time is discretized as $\underaccent{\tilde}{t}_n = n\,d\underaccent{\tilde}{t}$, where $d\underaccent{\tilde}{t}\ll\tau$ is the finite timestep, and we write $d\underaccent{\tilde}{\mathbf{u}}=\underaccent{\tilde}{\mathbf{u}}(\underaccent{\tilde}{t}_{n+1})-\underaccent{\tilde}{\mathbf{u}}(\underaccent{\tilde}{t}_n) = \underaccent{\tilde}{\mathbf{u}}_{n+1}-\underaccent{\tilde}{\mathbf{u}}_n$ such that the Euler-Langevin equations (\ref{eq:LangevinU})-(\ref{eq:LangevinN}) become
\begin{align}
d\underaccent{\tilde}{\mathbf{u}} &= - g_u \underaccent{\tilde}{\mathbf{u}}\,d\underaccent{\tilde}{t} + a_q \mathbf{n}_q\,d\underaccent{\tilde}{t} + a_\omega (\mathbf{n}_p \times \underaccent{\tilde}{\boldsymbol{\omega}})\,d\underaccent{\tilde}{t} +\sqrt{G_u d\underaccent{\tilde}{t}} \,\boldsymbol{{\cal N}}_u \label{eq:LangevinUNDim}\\
d\underaccent{\tilde}{\boldsymbol{\omega}} &=-g_\omega \underaccent{\tilde}{\boldsymbol{\omega}}\,d\underaccent{\tilde}{t} + b_q (\mathbf{n}_p\times \mathbf{n}_q)\,d\underaccent{\tilde}{t} + b_u (\mathbf{n}_p\times\underaccent{\tilde}{\mathbf{u}})\,d\underaccent{\tilde}{t} +\sqrt{G_\omega d\underaccent{\tilde}{t}}\, \boldsymbol{{\cal N}}_\omega \label{eq:LangevinOmegaNDim}\\
d\mathbf{n}_p &=\underaccent{\tilde}{\boldsymbol{\omega}}\times\mathbf{n}_p\,d\underaccent{\tilde}{t} \label{eq:LangevinNDim}
\end{align}
where all symbols on the right-hand side are evaluated at timestep $\underaccent{\tilde}{t}_n$. $\boldsymbol{{\cal N}}_u$ and $\boldsymbol{{\cal N}}_\omega$ are vectors containing independent normally distributed random variables with mean 0 and variance 1 at each timestep. For $T_p=T$ the parameters become
\begin{align}
g_u &= \tfrac{4}{3}(1+\tfrac{\pi}{16}(a^+ + a^-)),  & a_q &= \tfrac{4}{3}\,{\cal F}, & a_\omega & = \tfrac{1}{4}(a^+ - a^-), & G_u & = 2g_u {\cal G}, \\
g_\omega &=\tfrac{5}{2}\,\tfrac{1}{3}(a^+ + a^-), & b_q &=\tfrac{5}{2}\,\tfrac{1}{4}{\cal F}(a^+ - a^-), & b_u & = -\tfrac{5}{2}\,\tfrac{1}{4}(a^+ - a^-),
& G_\omega &= \tfrac{5}{2}\,2g_\omega {\cal G},
\end{align}
with
\begin{align}
{\cal F} = \frac{|\mathbf{q}|}{5p}\frac{1}{U_0}, \qquad
{\cal G} = \frac{kT}{m_p U_0^2} = \frac{m_p/m}{8\pi(R^3 n)^2}.
\end{align}
The factor $5/2$ in the coefficients of the equation for the angular velocity derives from $m_pR^2/I_p$ for a solid sphere. With these parameters the drift velocity and potential depth, equation (\ref{eq:potential}), become
\begin{align}
\underaccent{\tilde}{u}_d = \frac{a_q}{g_u}=\frac{\cal{F}}{1+\tfrac{\pi}{16}(a^+ + a^-)}, \qquad \frac{B_q}{kT} = \frac{b_q}{G_\omega/(2g_\omega)} = \frac{1}{4}\frac{\cal{F}}{\cal{G}}(a^+ - a^-).
\end{align}

The simulations figure \ref{fig:Langevin} is based on were run from $\underaccent{\tilde}{t}$ = 0 to 600 with a timestep of $d\underaccent{\tilde}{t} = 10^{-4}$ and initial conditions $\underaccent{\tilde}{\mathbf{u}}(0)=\underaccent{\tilde}{\boldsymbol{\omega}}(0)=\mathbf{0}$, $\mathbf{n}_p(0)=-\mathbf{e}_z$. Sampling was done starting from time $\underaccent{\tilde}{t}_\text{s}=6/g_\omega$ on a total of 100 individual runs.

The particular choice of parameters used for figure \ref{fig:Langevin} correspond to $\mathbf{n}_q=-\mathbf{e}_z$, ${\cal F} \approx 7.966$ and ${\cal G} \approx 0.7306$.

%%%%%%%%%%%%%%%%%%%%%%%%%%%%%%%%%%%%%%%%%%%%%%%%%%%%%%%%
\section{Thermal timescales  \label{sec:thermalTimescales}}

For a nanoparticle of radius 10~nm - 500~nm with a thermal diffusivity $a_p=\lambda_p/(\rho_p C_p) \simeq 10^{-7}$~m$^2$/s the timescale for internal thermal equilibration, $\tau_{\text{th},\text{int}} = R^2/a_p$, is 1~ns - 2.5~$\upmu$s.

We estimate the net heat flux at large Kn to the surface of the particle whose temperature is $\Delta T$ above the surroundings as $q\simeq 2\nu k\,\Delta T$. Then the timescale for thermal equilibration due to the external heat flux is $\tau_{\text{th}, \text{ext}} \simeq (\rho_P C_p R)/(3\nu k(a^++a^-) )$, or in units of the timescale used in appendix \ref{sec:nondimEuler}, $\tau_{\text{th},\text{ext}}/\tau \simeq C_pm/(k(a^++a^-))$. For $C_p$ = 1000~J/kg/K, we obtain $\tau_{\text{th},\text{ext}} /\tau \simeq 4.8$.

In the example of figure \ref{fig:Langevin} we had $\tau\approx 150\;\upmu$s as the timescale for reaching equilibrium of rotation and translation. Thus $\tau_{\text{th},\text{ext}} > \tau \gg \tau_{\text{th},\text{int}}$.

%%%%%%%%%%%%%%%%%%%%%%%%%%%%%%%

\end{document}